\documentclass[usenatbib]{mnras}

\usepackage{breakurl}

\usepackage{aas_macros}
\usepackage{graphicx}
\usepackage{latexsym}
\usepackage{amsfonts,amsmath,amssymb}
\usepackage{url}
\usepackage[utf8]{inputenc}
\usepackage[T1]{fontenc}
\usepackage{longtable}
\usepackage{multirow,booktabs}
\usepackage{txfonts}
\usepackage{xtab}
\usepackage{fp}
\usepackage{booktabs}

\newcommand{\es}  {{erg s$^{-1}$}}

\newcommand{\HST}{{\em Hubble Space Telescope}}
\newcommand{\HSTt}{{\em HST}}

\newcommand{\Chandra}{{\em Chandra}}

\newcommand{\msun}{M$_{\odot}$}
\newcommand{\nrgb}  {{$N_{\rm{RGB}}$}}
\newcommand{\rgbf}  {{$RGB_{\rm{frac}}$}}

\newcommand{\mv}  {{$M_{V}$}}
\newcommand{\lf}  {{$L_{\rm{frac}}$}}
\newcommand{\lx}  {{$L_{\rm{X}}$}}
\newcommand{\rh}  {{$r_{h}$}}

\defcitealias{nataf04-13}{N13}
\defcitealias{harris10-96}{H10}

\begin{document}

\title[GC-LMXB Metallicity]{Milky Way Globular Cluster Metallicity and Low-Mass X-ray Binaries: The Red Giant Influence}
\author[N. Vulic, P. Barmby, \& S. C. Gallagher]{N. Vulic$^1$\thanks{nvulic@uwo.ca}, P. Barmby$^1$, \& S. C. Gallagher$^1$  \\
$^1$Department of Physics \& Astronomy, Western University, London, ON, N6A 3K7, Canada}
\date{Received; Accepted}
\maketitle
\label{firstpage}
\pubyear{2017}

\begin{abstract}	

Galactic and extragalactic studies have shown that metal-rich globular clusters (GCs) are approximately three times more likely to host bright low-mass X-ray binaries (LMXBs) than metal poor GCs. There is no satisfactory explanation for this metallicity effect. We tested the hypothesis that the number density of red giant branch (RGB) stars is larger in metal-rich GCs, and thus potentially the cause of the metallicity effect. Using \HST\ photometry for 109 unique Milky Way GCs, we investigated whether RGB star density was correlated with GC metallicity. Isochrone fitting was used to calculate the number of RGB stars, which were normalized by the GC mass and fraction of observed GC luminosity, and determined density using the volume at the half-light radius (\rh). The RGB star number density was weakly correlated with metallicity [Fe/H], giving Spearman and Kendall Rank test $p$-values of 0.00016 and 0.00021 and coefficients $r_{s} = 0.35$ and $\tau = 0.24$ respectively. This correlation may be biased by a possible dependence of \rh\ on [Fe/H], although studies have shown that \rh\ is correlated with Galactocentric distance and independent of [Fe/H]. The dynamical origin of the \rh-metallicity correlation (tidal stripping) suggests that metal-rich GCs may have had more active dynamical histories, which would promote LMXB formation. No correlation between the RGB star number density and metallicity was found when using only the GCs that hosted quiescent LMXBs. A complete census of quiescent LMXBs in our Galaxy is needed to further probe the metallicity effect, which will be possible with the upcoming launch of $\emph{eROSITA}$.

\end{abstract}

\section{Introduction} \label{sec:intro}

Low-mass X-ray binaries (LMXBs) have companion stars of masses $\lesssim1.5$ \msun\ and are found in the field of a galaxy as well as in globular clusters (GCs). LMXBs form more efficiently in GCs due to increased stellar densities \citep{katz02-75, clark08-75, fabian08-75, pooley07-03}, and studies of Milky Way LMXBs found that their formation rate per unit stellar mass is 100 times greater in GCs compared to the field \citep{katz02-75, clark08-75}. Similar results have been found in elliptical galaxies \citep{sarazin10-03, jordan09-04, kim08-06, sivakoff05-07, kundu06-07, humphrey12-08, kim09-09}. 
Numerous relationships exist between GC properties and LMXBs. Globular clusters that are brighter/more massive, more compact (with smaller core radius $r_{c}$), and more metal-rich (redder) favour LMXB formation in both the Milky Way \citep{grindlay01-93, bellazzini02-95} and other nearby galaxies \citep{kundu07-02, maccarone05-04, jordan10-04, trudolyubov12-04, sivakoff05-07, kundu06-07, peacock10-10, paolillo08-11, kim02-13, agar11-13, mineo01-14, vulic08-14}. These studies have confirmed that metal-rich clusters are $\sim3$ times more likely to host LMXBs with limiting luminosities $>10^{36}$ \es\ for Milky Way and extragalactic observations. The dependence on mass and compactness is straightforward to explain because more stars and a higher density promote stellar interactions that create binaries. However, the metallicity dependence is still a mystery.

Various explanations have been suggested to explain the metallicity dependence, such as magnetic braking in main sequence stars \citep{ivanova01-06} or irradiation-induced stellar winds in low-metallicity stars \citep{maccarone05-04}. \citet{bellazzini02-95} was the first to indicate that the larger radii and masses of metal-rich stars would increase the rate of tidal capture in metal-rich GCs, thus increasing the number of LMXBs. \citet{ivanova12-12} posited that the difference in number densities and average masses of red giant stars in metal-rich versus metal-poor extragalactic GCs can explain the difference. Depending on their evolutionary state, GC-LMXBs can have either a main sequence, red giant, or white dwarf companion. Red giants promote dynamical formation of LMXBs via binary exchange interactions and physical collisions, serving as the seeds for dynamical formation of bright LMXBs. A strong argument for red giants being donor stars for neutron star GC-LMXBs with X-ray luminosities $>10^{37}$ \es\ is that these systems require companions with higher mass-loss rates. A full population synthesis study is still needed to confirm the red giant scenario, but here we attempt to address the observational effect.

We will use observations of Galactic GCs to compare the number and number density of red giants in clusters with and without LMXBs. In the Milky Way, 18 bright LMXBs are known in 14 GCs \citep{bahramian01-14}. Simulations by \citet{ivanova05-08} have shown that RGB stars are not expected to be donors for most Galactic GC-LMXB systems. However, RGB stars cannot be resolved in extragalactic GC cores, and so we are limited to the Milky Way population. While the prediction of \citet{ivanova12-12} was for bright ($>10^{37}$ \es) extragalactic neutron star GC-LMXBs, red giants serve as the seeds for bright ultracompact white dwarf-neutron star systems. In addition, main sequence-neutron star binaries (e.\ g.\ that are not X-ray sources) or LMXBs (below the `bright' limit) can evolve into the typical bright persistent LMXBs as observed in extragalactic studies.
To assess whether the number of red giants in LMXB-hosting clusters is proportionally larger than in GCs without LMXBs, we will use the method devised by \citet[][hereafter N13]{nataf04-13}. \citetalias{nataf04-13} used 72 Galactic GCs to study the red giant branch bump brightness and number counts by combining data from the \HST\ (\HSTt) Advanced Camera for Surveys (ACS) and Wide-field Planetary Camera 2 (WFPC2) instruments. They found that the `bump' brightness and number have a strong dependence on metallicity, foreshadowing a likely dependence of GC-LMXBs on the number of red giants. They reported the total number of red giants in 48 GCs, which we will use in addition to 61 other GCs to investigate the effect of red giants on the metallicity dependence of LMXBs in Galactic GCs.

\section{Data} \label{sec:data}

We use data from two large \HSTt\ surveys of Milky Way GCs. The first was carried out by \citet{piotto09-02} using the F439W and F555W filters on the WFPC2 instrument (WF2/WF3/WF4 each has a resolution of 0.1\arcsec\ pixel$^{-1}$, PC1 has a resolution of 0.046\arcsec\ pixel$^{-1}$). They studied 74 GCs with a wide range of properties by investigating colour-magnitude diagrams (CMDs), which are complete to approximately the main sequence turnoff. The Planetary Camera was centred on the cluster centre in each case. The more recent treasury survey by \citet{sarajedini04-07} and \citet{dotter07-07} used the F606W and F814W filters on the ACS Wide Field Camera instrument (resolution of 0.05\arcsec\ pixel$^{-1}$). Each cluster was centred in the ACS field. The program studied 71 GCs and obtained photometry with $S/N\gtrsim10$ for stars down to 0.2 \msun.
The benefit of using these treasury surveys is that in studying various aspects of Galactic GCs the authors produced precise and consistent photometric catalogues. Both catalogues have carried out artificial star tests that confirm completeness on the RGB. \citetalias{nataf04-13} used 72 Galactic GCs by combining the ACS and WFPC2 surveys. Because the authors were studying the RGB bump, they only chose clusters that had well-populated RGBs and RGB bumps ($N_{\rm{RGBB}}\geq10$), and were not affected by differential reddening. They called this their `gold' sample, which consisted of 48 GCs for which they reported RGB numbers. The remaining 24 GCs that made up their `silver' sample had anomalous RGB bumps.
The RGB numbers from the 48 GCs in the gold sample of \citetalias{nataf04-13} will be used in our study; we independently determine \nrgb\ for the remaining clusters. Combining the ACS and WFPC2 datasets there are 109 unique Galactic GCs, of which \citetalias{nataf04-13} analysed 48. The remaining 61 GCs consist of 34 from the ACS survey and 27 from the WFPC2 survey. To be consistent in our analysis we followed the methods of \citetalias{nataf04-13} to obtain RGB numbers for these GCs. First, we summarize some issues regarding the data that have been addressed by \citetalias{nataf04-13}. The photometric filters used in each survey were different, and thus a standard calibration needed to be adopted. Photometric values were transformed from the F439W/F555W and F606W/F814W filters for WFPC2 and ACS, respectively, into the Johnson ($V,B-V$) and ($I,V-I$) planes in the original catalogue papers. A comparison between 13 GCs that were common to both surveys found that the difference in the derived $V$ magnitudes was negligible. The magnitudes from both catalogues are not reddening-corrected. Using both ACS and WFPC2 data it is possible that crowding could have been an issue. However, because PC1 was centred on the core of each GC, this effect will be reduced given the similar resolution (to ACS) and concentration of RGB stars in cluster cores.

\section{Isochrone Fitting} \label{sec:isc}

Before we determined the number of RGB stars, we plotted isochrones for each of the 61 GCs to guide our analysis. We plotted isochrones only to identify the different regions of the CMD, such as the subgiant, red giant, asymptotic giant, and horizontal branches. We do not aim to determine the age, distance, or metallicity of a GC using this method but only to better approximate the RGB.
We used the isochrones provided by the Dartmouth Stellar Evolution Database\footnote{http://stellar.dartmouth.edu/models}, which provides isochrone grids based on the original 2008 version photometric systems \citep{dotter09-08}. We used the ACS Galactic Globular Cluster Survey isochrones that include the BVI/F606W/F814W empirical colours for the ACS dataset. These isochrones were created by \citet{dotter07-07} specifically for the ACS catalogue we use here. We used the empirical BVI colour isochrones from \citet{vandenberg08-03} for our WFPC2 dataset.
All isochrone grids have age intervals of 250 Myr between ages of $1-5$ Gyr, and 500 Myr intervals between ages of $5-15$ Gyr. Metallicities for [Fe/H] range from $-2.5$ to $+0.5$, with steps of 0.5 in the range of interest for our work ($0.0$ to $-2.5$). The $\alpha$-enhancement [$\alpha$/Fe] is another probe of the metallicity, and in the models it refers to enhancements in the following $\alpha$-capture elements: O, Ne, Mg, Si, S, Ca, Ti. The value of [$\alpha$/Fe] ranges from $-2$ to 8 in steps of 2. The models assume an initial He mass fraction Y${\rm{init}}$ = 0.245 + 1.54Z, with additional grids using Y$_{\rm{init}}$  of 0.33 and 0.40 for [$\alpha$/Fe] of 0.0 and 0.4 (see Table 2 of \citet{dotter07-07} for more details).

To determine the best-fitting isochrones for each of our GCs, we first assumed the metallicity [Fe/H] given by \citet[][2010 edition, hereafter H10]{harris10-96}\footnote{http://www.physics.mcmaster.ca/Globular.html} and found the nearest value of [Fe/H] from the isochrone grid. We then plotted isochrones with ages ranging from 1$-$15 Gyr for each of the 10 different combinations of [$\alpha$/Fe] and initial He mass fraction values on our CMDs. The isochrone $BVI$ magnitudes were adjusted for each cluster using its distance from the Sun \citepalias{harris10-96}. Because we used the raw Johnson $BVI$ magnitudes that were not corrected for reddening, we shifted our isochrones based on the $E(B-V)$ values given in \citetalias{harris10-96}. For the ACS survey, since we were working in the ($I, V-I$) plane, we converted the reddening using $E(V-I) = 1.26\times E(B-V)$ \citep{cardelli10-89, barmby02-00}. We obtained $A_{V}$ and $A_{I}$ using the standard Galactic extinction law \citep{cardelli10-89}. A number of GCs had all their isochrones shifted away from the main sequence and RGB, with no overlap. This arose from the uncertainty in the distance to a GC and also the conversion factor for reddening, which caused inaccurate $E(V-I)$ values.
For these isochrones we shifted the $E(V-I)$ values to account for this effect.
We chose the best-fitting CMD that most accurately represented the main sequence, subgiant and red giant branches. We cross-checked the ages determined from isochrone fitting with results from \citet{dotter01-10}, who used the \HSTt\ magnitude plane for isochrone fitting, to ensure our results were consistent. We report the results of our isochrone fitting in Table \ref{tab:isc}, where the parameters do not reflect precise values for each cluster but instead the values for a specific isochrone.

\begin{table}
\caption{Isochrone Fitting Parameters\label{tab:isc}}
\resizebox{\columnwidth}{!}{%

\begin{tabular}{@{}c c  c  c  c  c  c  c@{}}
\hline\hline
Globular Cluster	&	Age (Gyr)	&	[Fe/H]	&	[$\alpha$/H]	&	E(B$-$V)	&	E(B$-$V) Shift	&	E(V$-$I)	&	Instrument	\\
\hline
        Arp 2 & 13.0 & -2.0 & 0.2 & 0.15 & 0.05 & 0.19 & ACS
 \\ E 3 & 13.0 & -1.0 & 0.2 & 0.33 & 0.03 & 0.42 & ACS
 \\ IC 4499 & 12.0 & -1.5 & 0.2 & 0.25 & 0.02 & 0.32 & ACS
 \\ NGC 288 & 12.5 & -1.5 & 0.4 & 0.03 & 0.00 & 0.04 & ACS
 \\ NGC 2298 & 13.0 & -2.0 & 0.2 & 0.27 & 0.12 & 0.33 & ACS
 \\ NGC 4147 & 13.0 & -2.0 & 0.2 & 0.06 & 0.04 & 0.08 & ACS
 \\ NGC 4590 & 13.0 & -2.0 & 0.2 & 0.08 & 0.03 & 0.09 & ACS
 \\ NGC 4833 & 13.0 & -2.0 & 0.4 & 0.38 & 0.06 & 0.48 & ACS
 \\ NGC 5053 & 13.5 & -2.5 & 0.2 & 0.04 & 0.03 & 0.05 & ACS
 \\ NGC 5139 & 12.0 & -1.5 & 0.2 & 0.17 & 0.05 & 0.21 & ACS
 \\ NGC 5466 & 13.0 & -2.0 & 0.2 & 0.04 & 0.04 & 0.04 & ACS
 \\ NGC 6101 & 13.0 & -2.0 & 0.2 & 0.15 & 0.10 & 0.19 & ACS
 \\ NGC 6121 & 12.5 & -1.0 & 0.4 & 0.41 & 0.07 & 0.52 & ACS
 \\ NGC 6144 & 13.5 & -2.0 & 0.2 & 0.49 & 0.13 & 0.62 & ACS
 \\ NGC 6366 & 12.0 & -0.5 & 0.2 & 0.71 & 0.00 & 0.89 & ACS
 \\ NGC 6397 & 13.5 & -2.0 & 0.2 & 0.21 & 0.04 & 0.27 & ACS
 \\ NGC 6426 & 12.0 & -2.0 & 0.2 & 0.45 & 0.09 & 0.56 & ACS
 \\ NGC 6496 & 12.0 & -0.5 & 0.2 & 0.21 & 0.06 & 0.26 & ACS
 \\ NGC 6535 & 13.0 & -2.0 & 0.2 & 0.48 & 0.14 & 0.60 & ACS
 \\ NGC 6656 & 12.5 & -1.5 & 0.2 & 0.40 & 0.06 & 0.50 & ACS
 \\ NGC 6715 & 12.0 & -1.5 & 0.2 & 0.17 & 0.02 & 0.21 & ACS
 \\ NGC 6717 & 13.0 & -1.5 & 0.2 & 0.27 & 0.05 & 0.34 & ACS
 \\ NGC 6779 & 13.5 & -2.0 & 0.2 & 0.28 & 0.02 & 0.35 & ACS
 \\ NGC 6809 & 13.5 & -2.0 & 0.2 & 0.14 & 0.06 & 0.18 & ACS
 \\ NGC 6838 & 12.5 & -1.0 & 0.2 & 0.27 & 0.02 & 0.34 & ACS
 \\ NGC 7099 & 13.0 & -2.5 & 0.2 & 0.09 & 0.06 & 0.11 & ACS
 \\ Palomar 1 & 7.0 & -0.5 & 0.2 & 0.16 & 0.01 & 0.20 & ACS
 \\ Palomar 12 & 9.5 & -1.0 & 0.0 & 0.06 & 0.04 & 0.08 & ACS
 \\ Palomar 15 & 13.0 & -2.0 & 0.2 & 0.47 & 0.07 & 0.59 & ACS
 \\ Palomar 2 & 12.0 & -1.5 & 0.2 & 1.19 & -0.05 & 1.50 & ACS
 \\ Pyxis & 12.0 & -1.0 & 0.2 & 0.28 & 0.07 & 0.35 & ACS
 \\ Ruprecht 106 & 10.0 & -1.5 & 0.2 & 0.23 & 0.03 & 0.28 & ACS
 \\ Terzan 7 & 8.0 & -0.5 & 0.0 & 0.08 & 0.01 & 0.10 & ACS
 \\ Terzan 8 & 13.0 & -2.0 & 0.4 & 0.16 & 0.04 & 0.20 & ACS
 \\ IC 1257 & 12.0 & -1.5 & 0.0 & 0.73 & - & - & WFPC2
 \\ NGC 1904 & 12.0 & -1.5 & 0.6 & 0.01 & - & - & WFPC2
 \\ NGC 2419 & 12.5 & -2.0 & 0.4 & 0.08 & - & - & WFPC2
 \\ NGC 4372 & 13.5 & -2.0 & 0.8 & 0.39 & - & - & WFPC2
 \\ NGC 5694 & 13.5 & -2.0 & 0.8 & 0.09 & - & - & WFPC2
 \\ NGC 5946 & 13.5 & -1.5 & 0.8 & 0.54 & - & - & WFPC2
 \\ NGC 6235 & 12.0 & -1.5 & 0.6 & 0.31 & - & - & WFPC2
 \\ NGC 6256 & 10.0 & -1.0 & 0.0 & 1.09 & - & - & WFPC2
 \\ NGC 6266 & 12.0 & -1.0 & 0.0 & 0.47 & - & - & WFPC2
 \\ NGC 6273 & 12.0 & -1.5 & 0.2 & 0.38 & - & - & WFPC2
 \\ NGC 6287 & 13.5 & -2.0 & 0.8 & 0.60 & - & - & WFPC2
 \\ NGC 6293 & 13.5 & -2.0 & 0.4 & 0.36 & - & - & WFPC2
 \\ NGC 6316 & 13.5 & -0.5 & 0.0 & 0.54 & - & - & WFPC2
 \\ NGC 6325 & 13.5 & -1.5 & 0.6 & 0.91 & - & - & WFPC2
 \\ NGC 6342 & 13.5 & -0.5 & 0.0 & 0.46 & - & - & WFPC2
 \\ NGC 6355 & 13.5 & -1.5 & 0.2 & 0.77 & - & - & WFPC2
 \\ NGC 6380 & 5.0 & -1.0 & 0.8 & 1.17 & - & - & WFPC2
 \\ NGC 6401 & 12.5 & -1.0 & 0.6 & 0.72 & - & - & WFPC2
 \\ NGC 6440 & 11.0 & -0.5 & 0.0 & 1.07 & - & - & WFPC2
 \\ NGC 6453 & 13.5 & -1.5 & 0.0 & 0.64 & - & - & WFPC2
 \\ NGC 6517 & 9.0 & -1.0 & 0.0 & 1.08 & - & - & WFPC2
 \\ NGC 6522 & 13.0 & -1.5 & 0.8 & 0.48 & - & - & WFPC2
 \\ NGC 6539 & 7.0 & -0.5 & 0.0 & 1.02 & - & - & WFPC2
 \\ NGC 6540 & 12.0 & -1.5 & 0.0 & 0.66 & - & - & WFPC2
 \\ NGC 6544 & 13.0 & -1.5 & 0.2 & 0.76 & - & - & WFPC2
 \\ NGC 6642 & 13.5 & -1.5 & 0.8 & 0.40 & - & - & WFPC2
 \\ NGC 6712 & 12.5 & -1.0 & 0.2 & 0.45 & - & - & WFPC2
 \\ 
\hline
\end{tabular}
}
\begin{list}{}{}
\item Isochrone fitting parameters for the 61 GCs without reported RGB numbers in \citetalias{nataf04-13}. The age, metallicity ([Fe/H]), and helium enhancement ([$\alpha$/Fe]) are the values used for the best-fitting isochrones and do not reflect precise values for each cluster. The $E(B-V)$ values are taken from \citetalias{harris10-96}, while the $E(V-I)$ values were derived using $E(V-I) = 1.26\times E(B-V)$ \citep{cardelli10-89, barmby02-00}. $E(V-I)$ values were only needed for ACS data because CMDs were created in the ($I, V-I$) plane. This conversion was not accurate for all clusters and therefore an offset in the original $E(B-V$) values for ACS clusters was introduced to properly align the isochrone. This value is shown in the sixth column as $E(B-V)$ Shift. The last column indicates which camera was used to observe the cluster.
\end{list}
\end{table}

\section{Red Giant Branch Star Numbers} \label{sec:rgbfit}

\begin{figure*}
\includegraphics[]{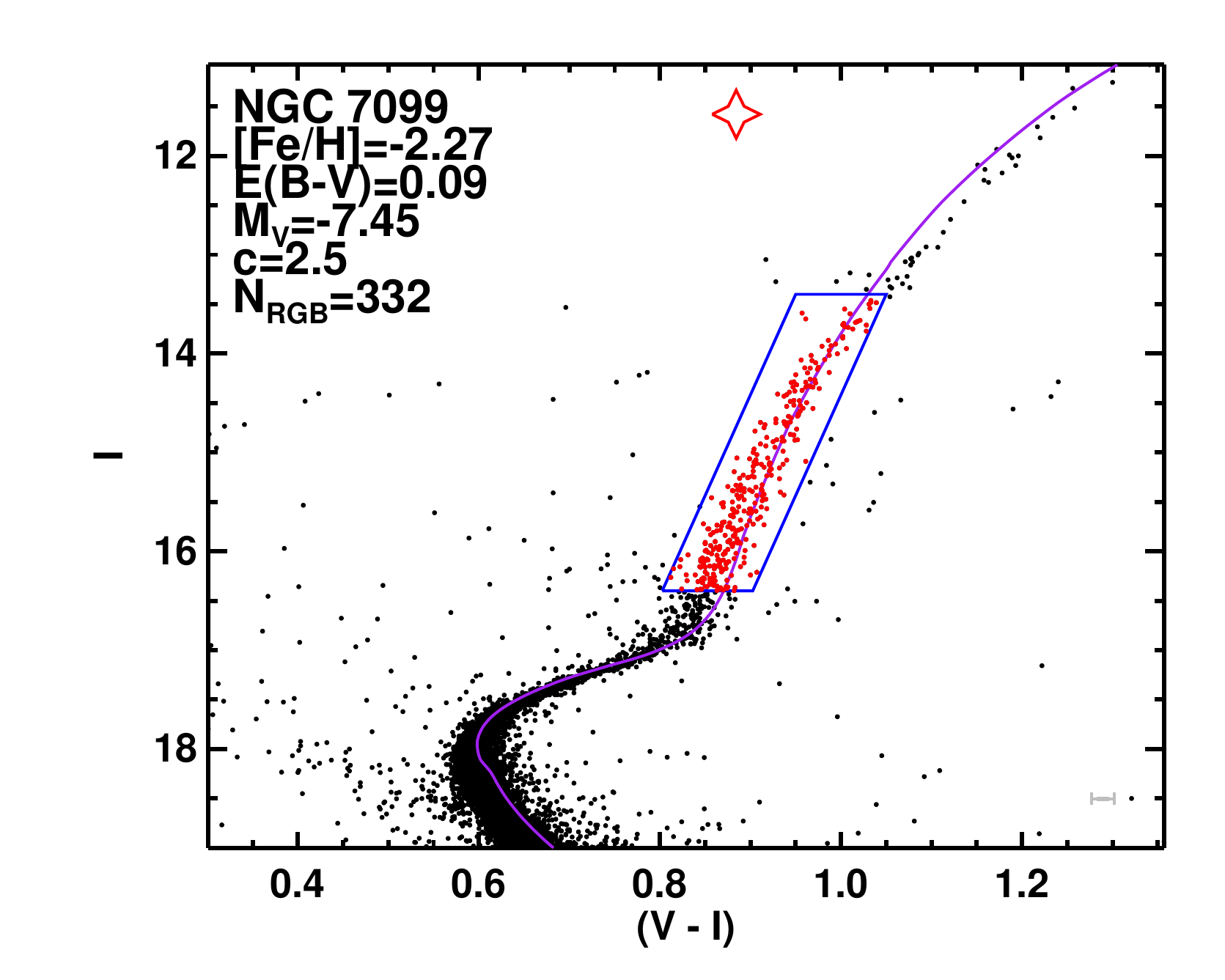}
\caption{Colour-magnitude diagram of NGC 7099 in the Johnson $(I,V-I)$ plane without reddening correction. The overplotted best-fitting isochrone (purple curve) has parameters given in Table \ref{tab:isc}. The blue parallelogram corresponds to the conservative estimate of the region where red giant branch stars exist, giving the number of red giant branch stars in the cluster \nrgb. The open star at top-centre indicates that the cluster hosts a quiescent LMXB. Various cluster parameters from \citetalias{harris10-96}, including the absolute magnitude \mv\ and concentration parameter $c$ are shown at top-left. Average photometric uncertainties are represented by the grey error bar at bottom-right. The CMDs for the remaining ACS and WFPC2 GCs are in Appendices \ref{app:cmds-acs} and \ref{app:cmds-wfpc2}, respectively.}\label{fig:cmdn7099}
\end{figure*}

In order to determine the number of RGB stars in each GC, \nrgb, we inspected the CMDs of each cluster individually to create a bounded area that represented the RGB. Following the technique of \citetalias{nataf04-13} (D.\ M.\ Nataf, priv.\ comm.), we required the bounding region for the RGB to be as long and wide as possible. We selected the lower bound to be approximately 0.5 magnitudes above the mean brightness of the subgiant branch. This attempts to eliminate any foreground stars that may be contaminating the region. Specifically, some clusters had abrupt subgiant branches where the main sequence and RGB are not well-separated, introducing more contamination from foreground stars near the subgiant branch region of the CMD. The upper bound was selected based on the decreasing density of stars towards the tip of the RGB, and the beginning of the curve in the distribution of stars that represents the start of the asymptotic giant branch. In some cases the horizontal branch could be used as a reference point for the end of the RGB. We limited the width of the parallelogram resulting from the lower and upper bounds for two reasons. First, moving too far blueward of the RGB can result in horizontal branch stars being included in our estimate of \nrgb. Second, the wider our parallelogram, the higher incidence of foreground stars we can expect to contaminate our sample, further biasing our approximation. In Figure \ref{fig:cmdn7099} we show the CMD of NGC 7099, a metal-poor GC with minimal extinction and large concentration parameter. The blue parallelogram reflects the criteria just summarized, and includes 332 RGB stars. The CMDs for the remaining ACS and WFPC2 GCs are in Appendices \ref{app:cmds-acs} and \ref{app:cmds-wfpc2}, respectively. In Table \ref{tab:rgb-values} we present the \nrgb\ values and the properties for each GC in our study. This includes cluster parameters from \citetalias{harris10-96}: absolute V magnitude \mv, metallicity [Fe/H], concentration parameter ($log\frac{r_{\rm{tidal}}}{r_{\rm{core}}}$), core radius $r_{\rm{c}}$, half-light radius $r_{\rm{h}}$, and the distance from the Sun $R_{\rm{Sun}}$. The upper and lower limits (UL and LL) on \nrgb\ were derived using $\sqrt{N}$ for values $\ge50$ and Poisson statistics \citep{gehrels04-86} for values $<50$. We also include the mass and mass-to-light ratios $M/L$ taken from Table 8 (column 6) of \citet[][see Section \ref{sec:gcmass-norm} for more details]{mclaughlin12-05}. Masses were derived by multiplying  $M/L$ by total cluster luminosity derived from \mv.

\section{Normalizations} \label{sec:norm}

\subsection{Globular Cluster Luminosity and Foreground Contamination} \label{sec:lumfgstarcont}

A number of normalizations were required in order to effectively use the \nrgb\ parameter in our analysis. The first and most obvious is for the instrument field of view. For the 109 unique GCs for which \nrgb\ was determined, 71 come from the ACS survey and 38 from the WFPC2 survey. The ACS instrument has a slight parallelogram shape and is comprised of the Wide Field Camera 1 and 2, with a total field of view of 202\arcsec\ by 202\arcsec. The WFPC2 is made up of 4 CCDs, 3 wide field cameras (WF2, WF3, WF4), each with field of view 75\arcsec\ by 75\arcsec, and the Planetary Camera (PC1), having a 32\arcsec\ by 32\arcsec\ field of view. Therefore the WFPC2 is essentially a square of area 150\arcsec\ a side with a piece missing due to the size of PC1. We needed to account for the difference in observed area of each cluster including the different fields of view of ACS and WFPC2. All GCs in the ACS survey were observed twice in each filter, with the fields overlapping each other close to the 100\% level. The WFPC2 survey was a snapshot program where the numerous exposures overlapped similar to the ACS survey. We determined the total area observed for each cluster (in pc$^{2}$) by counting the nonzero pixels in the merged images for both the ACS and WFPC2 surveys. 

However, the issue was exacerbated because the fraction of each GC observed (physical size) depends on the instrument field of view and the cluster's distance. Therefore the appropriate correction involved determining the fraction of cluster luminosity observed, \lf$=L_{\rm{obs}}/L_{\rm{total}}$, where $L_{\rm{obs}}$ is observed cluster luminosity and $L_{\rm{total}}$ is the total luminosity of the cluster. $L_{\rm{total}}$ was calculated by using GC absolute magnitudes\footnote{These absolute magnitudes are corrected for foreground Galactic extinction.} \mv\ from \citetalias{harris10-96}. No uncertainties are reported for any of the $M_{V}$ values and so we adopt a conservative universal uncertainty of 0.2 magnitudes, consistent with the uncertainty of the globular cluster luminosity function peak \citep{kavelaars03-97}.
For this calculation and others, we used the reference values from \citet{mamajek08-12} for the Sun\footnote{\url{https://sites.google.com/site/mamajeksstarnotes/basic-astronomical-data-for-the-sun}} of L$_{\odot,\rm{bol}} = 3.8270 \pm 0.0014 \times10^{33}$ \es, M$_{\odot,\rm{bol}} = 4.7554 \pm 0.0004$, $m_{\odot,\rm{bol}} =  -26.8167 \pm 0.0004$. 
The observed cluster luminosities, $L_{\rm{obs}}$, were calculated by summing the photometry from each cluster field. Because each star in a field of view has an apparent magnitude $m_{V}$ associated with it, we converted to flux and determined $L_{\rm{obs}}$. By working in the $V-$band for all clusters and using $M_{V}$ to get $L_{\rm{total}}$ we remain consistent within one passband. However, by summing the flux from all stars in a given field of view we were also including the contribution from foreground stars, which biases our calculations. This has the opposite effect of extinction in that it would raise our total observed luminosity. To account for this effect we estimated the contamination from Galactic foreground stars using the Besan\c{c}on models\footnote{\url{http://model.obs-besancon.fr/}} \citep{robin10-03}.
Using the (Galactic) co-ordinates from the centres of the ACS and WFPC2 images (not always the cluster centres), along with the solid angle in deg$^{2}$, we obtained a detailed list of parameters of foreground stars for each cluster. For the solid angle, we used the area for each cluster field of view as described above. The models produce reliable predictions for the luminosity and colour distributions in the optical/near-infrared for the stars expected to be in the field of view. We used the apparent magnitudes $m_{V}$ and given visual extinctions $A_{V}$ from the model to determine the total absolute $V$ magnitude from foreground stars, \mv (fgstars). Subtracting this foreground flux from the total observed flux allowed us to calculate accurate observed cluster luminosities. $L_{\rm{total}}$ represents the actual cluster luminosity after removing the foreground star contamination and correcting for extinction.

\begin{figure*}
\includegraphics[]{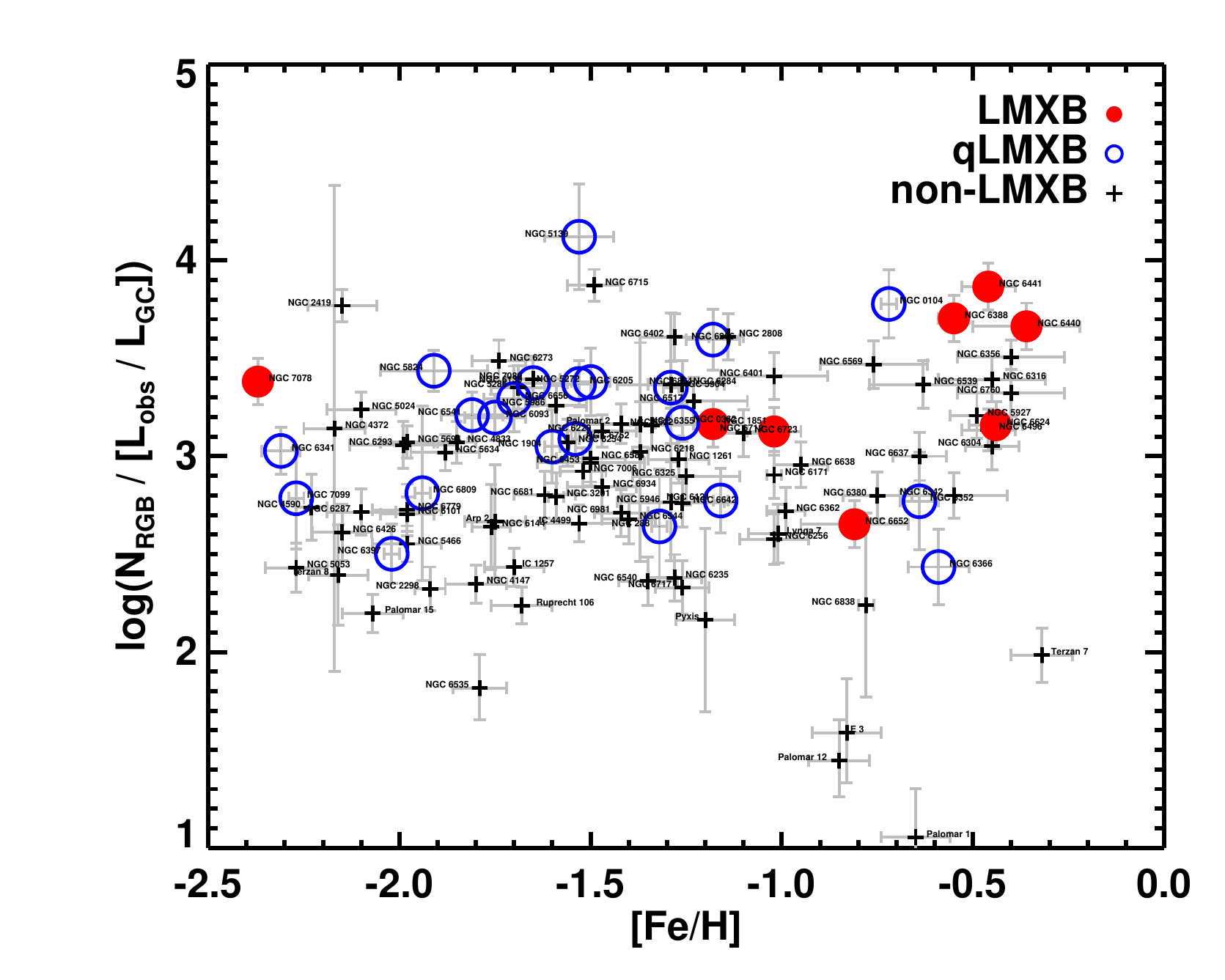}
\caption{RGB star number vs. metallicity [Fe/H]. Black plusses represent normal GCs, open blue circles are GCs hosting quiescent LMXBs and filled red circles are GCs hosting bright LMXBs. NGC 6440 and NGC 7078 each have 2 LMXBs. GC names are indicated near each datapoint. The \nrgb\ parameter is expected to scale linearly with mass and thus needs to be normalized by mass in order to assess any potential relationship between RGB stars and LMXBs.}\label{fig:rgb-lfrac}
\end{figure*}

For some of the clusters we were not able to accurately determine a value for \lf, likely due to the uncertainty associated with distance, extinction, and the foreground star modelling/photometry. For 3 GCs, namely Lynga 7, Palomar 12, and Terzan 7, we had to remove the brightest star from the photometric catalogue. These stars were $2-3$ magnitudes brighter than the next brightest star, which was part of the smooth distribution of cluster stars. For NGC 6453 and Palomar 2, we used updated extinction values $E(B-V)$ from \citet{schlafly08-11} obtained using the NASA/IPAC database\footnote{\url{http://irsa.ipac.caltech.edu/applications/DUST/}}. Our \lf\ values vary from $5\%-97\%$, where the lower limit corresponds to NGC 4372, a large, nearby (5.8 kpc) GC that was observed with the small field of view of WFPC2. The upper limit of 97\% is for Palomar 1, a very dense GC with small core and half-light radii observed with ACS. In Figure \ref{fig:rgb-lfrac} we show the number of RGB stars normalized by the observed GC luminosity vs. metallicity. Clusters with LMXBs are indicated by filled red circles and clusters with qLMXBs by open blue circles. NGC 6440 and NGC 7078 (M15) each have two LMXBs. From our sample of 109 GCs we have 10 bright LMXBs in 8 GCs. We used results compiled by \citet{verbunt04-06} and \citet{bahramian01-14} for LMXBs and qLMXBs in Galactic GCs (see Table \ref{tab:rgb-values}). The uncertainties on [Fe/H] were taken from \citet{carretta12-09}. For the five GCs in our sample that didn't have uncertainties on [Fe/H], IC 1257, Lynga 7, NGC 6426, NGC 6540, and Terzan 8, we set them to the mean uncertainty value of the remaining GCs. The \nrgb\ distribution still needs to be normalized in order to appropriately assess its impact on LMXB formation. We present the values of \lf, $\Delta$\lf, and \mv (fgstars) for each cluster in Table \ref{tab:rgb-values}.

\subsection{Globular Cluster Mass} \label{sec:gcmass-norm}

\begin{figure*}
\includegraphics[]{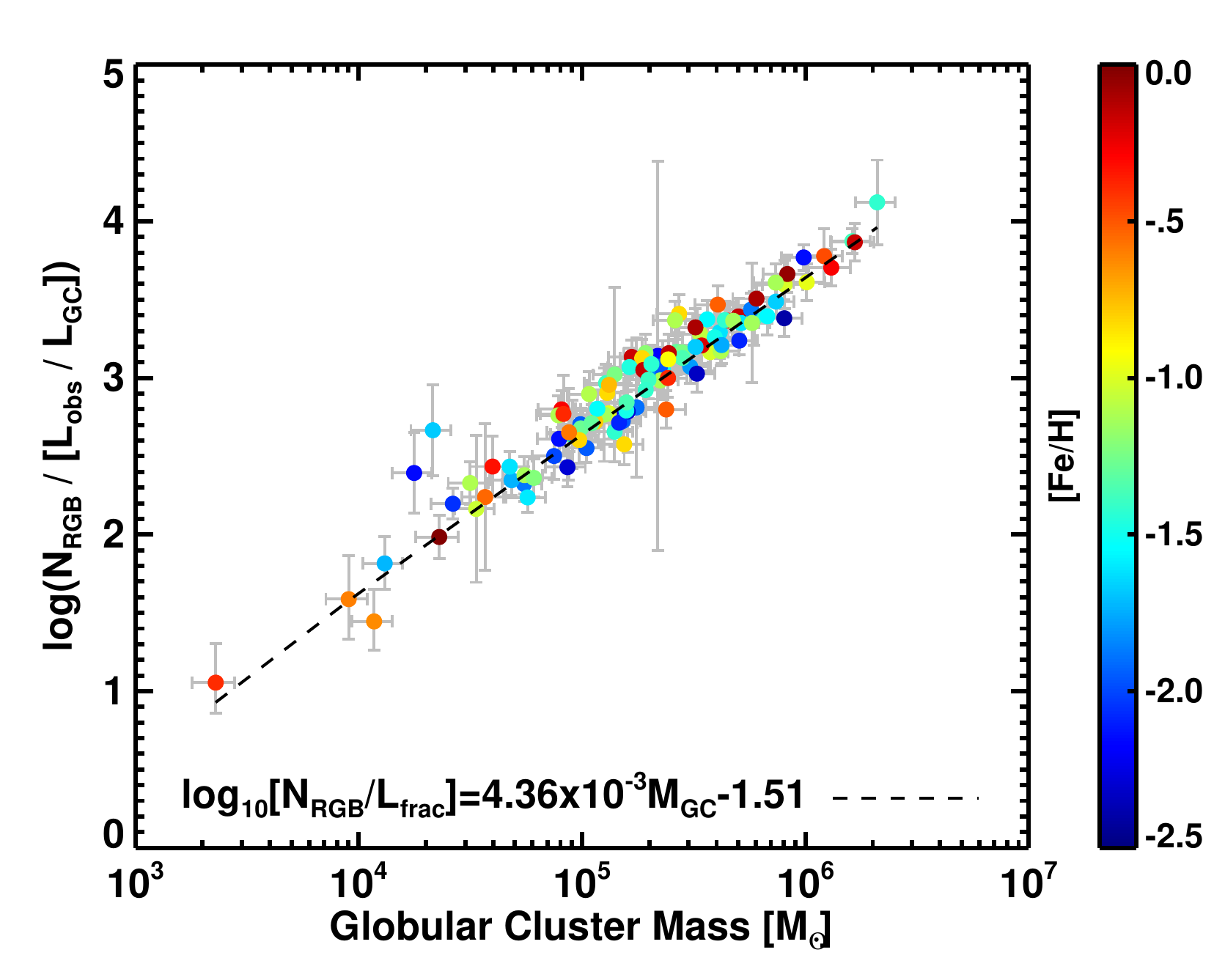}
\caption{The number of RGB stars \nrgb\ normalized by the fraction of observed cluster luminosity \lf\ vs. the cluster mass. GC metallicity [Fe/H] is indicated by the colourbar to the right. The strong correlation between the total number of RGB stars in a GC and its mass is evident, as expected. The dashed line shows the weighted least squares fit to the data, which is given in equation \ref{eq:rgbnum-mass}.}\label{fig:nrgb-gcmass}
\end{figure*}

Normalizing \nrgb\ by the fraction of observed luminosity accounts for RGB stars that would have been observed had the field of view been larger.
However, even if each GC had \lf$=1$, our result would still be biased towards clusters that are more massive, which by extension have a larger number of RGB stars. Because the relative \nrgb\ is important, we needed to normalize \nrgb\ by the mass of each GC. We calculated cluster masses using our value for the total cluster luminosity $L_{\rm{total}}$ and the mass-to-light ratio $M/L$ of the cluster. \citet{mclaughlin12-05} tabulated $V-$band $M/L$ (Table 8, column 6 of their paper) for 148 Galactic GCs, which includes the 109 GCs in our work (see Table \ref{tab:rgb-values}). The $M/L$ ratios were derived using the code from \citet{bruzual10-03} and the disc initial mass function of \citet{chabrier07-03}. We used our cluster luminosities to determine the mass of each GC in our sample. In Figure \ref{fig:nrgb-gcmass} we show the relationship between \nrgb/\lf\ and the GC mass. 

\begin{figure*}
\includegraphics[]{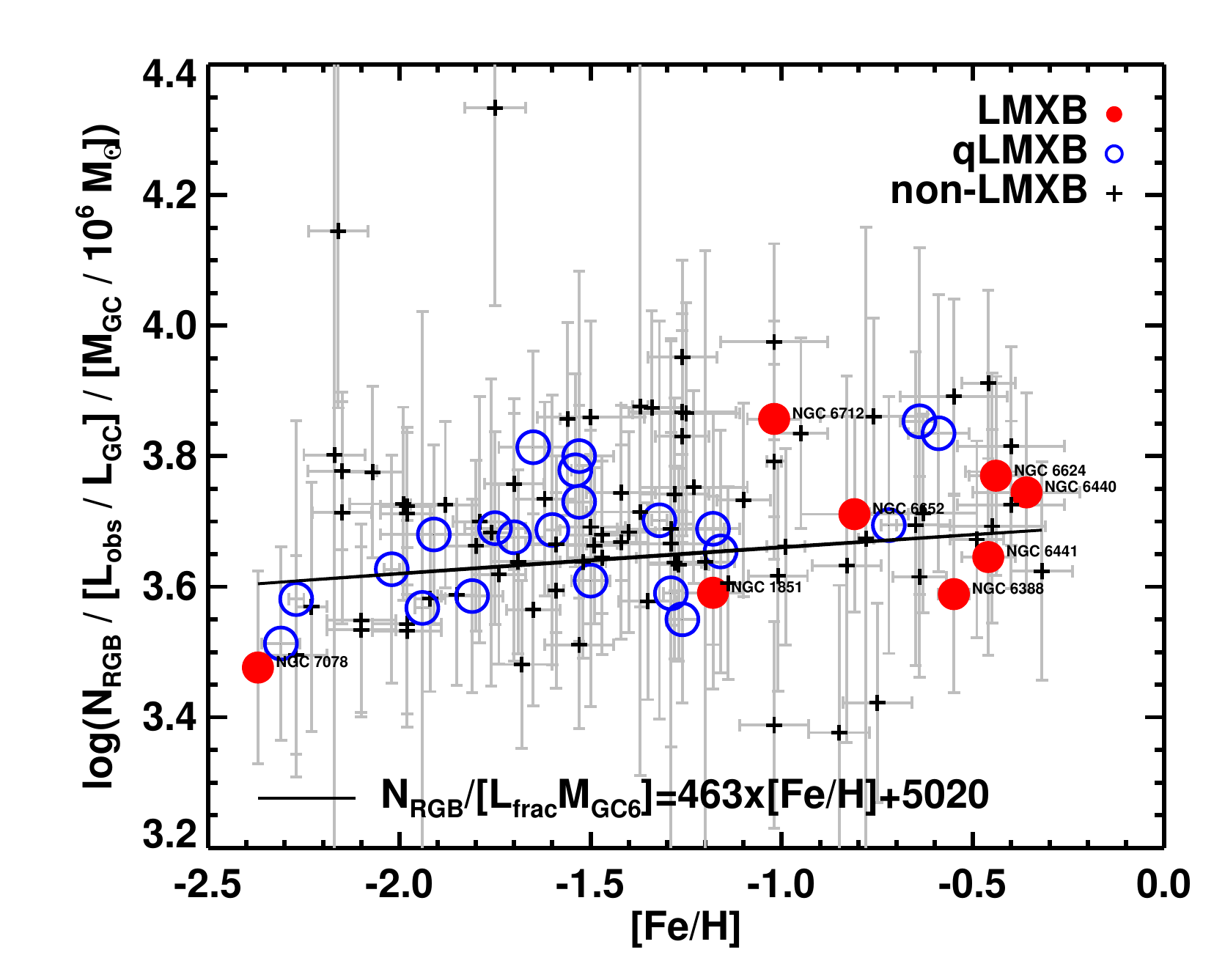}
\caption{\rgbf\ parameter (RGB star number normalized by GC mass) vs. metallicity [Fe/H]. Black plusses represent normal GCs, open blue circles are GCs hosting quiescent LMXBs and filled red circles are GCs hosting bright LMXBs. NGC 6440 and NGC 7078 each have 2 LMXBs. Names of GCs with LMXBs are indicated near each datapoint. The weighted least squares fit (equation \ref{eq:rgbnum-massmetal}) is shown, which is consistent with a flat distribution. This result indicates no relationship with the number of RGB stars per \msun\ and metallicity of a GC.}\label{fig:rgb-lfracmass}
\end{figure*}

As expected, the number of RGB stars increases with GC mass. The datapoint with a large uncertainty in \nrgb\ is NGC 4372, which happens to be the cluster for which \lf\ was 5\%, having a large uncertainty of 15\% (300\% relative). A weighted least-squares fit to the data produced the relationship in equation \ref{eq:rgbnum-mass}.
\begin{align}
\log_{10}\left[\frac{N_{\rm{RGB}}}{L_{\rm{frac}}}\right]	& = & (0.00436 \pm 0.00013)\times \frac{\rm{M}_{GC}}{\rm{M}_{\odot}} - (1.51 \pm 4.99)	\label{eq:rgbnum-mass}
\end{align}
In addition, GC mass has a strong influence on the presence of an LMXB \citep[e.g.][]{sivakoff05-07, vulic08-14}. Therefore not only are we removing the intrinsic dependence of RGB number on mass but also a parameter (mass, via \nrgb) that is known to promote the production of LMXBs. When normalizing by GC mass we obtain the number of RGB stars per unit mass, \nrgb\ \msun$^{-1}$, which we call \rgbf. In Figure \ref{fig:rgb-lfracmass} we plot \rgbf\ vs. metallicity using the same format as Figure \ref{fig:rgb-lfrac}.

Figure \ref{fig:rgb-lfracmass} shows that \rgbf\ varies very little with metallicity, where only a slightly positive correlation may exist. The much larger uncertainties in the \rgbf\ parameter are a result of the uncertainty in GC masses, which propagate from the values for total cluster luminosity $L_{\rm{total}}$. A weighted least squares fit to the data is shown by the solid line and reproduced in equation \ref{eq:rgbnum-massmetal}, where M$_{\rm{GC6}}$ is the GC mass in terms of $10^{6}$ \msun.
\begin{align}
RGB_{\rm{frac}} = \frac{N_{\rm{RGB}}}{L_{\rm{frac}}\rm{M}_{GC6}}		& = & \nonumber \\ (462.93 \pm 291.70)\times\rm{[Fe/H]} +  (5019.84 \pm 452.36)		\label{eq:rgbnum-massmetal}
\end{align}
We have effectively removed the bias that would exist in this relationship due to mass and thus have a quantity \rgbf\ that can be independently compared to the metallicity. This is important because a mass-metallicity relationship could also affect our results. In early-type galaxies, the brighter metal-poor GCs show a relationship between mass and metallicity, dubbed the `blue tilt', thought to arise from self-enrichment in GCs \citep{peng03-06, harris01-06, mieske12-06, strader12-06, bailin04-09}.
In the Milky Way, a mass-metallicity relationship has not been detected for several reasons (e.g.\ cluster to cluster scatter in mean [Fe/H], sample size, mass limit), although its existence has not been ruled out \citep{strader11-08}. Therefore it is beneficial to remove the dependence on mass to avoid any intrinsic dependence of massive metal-poor Galactic GCs on metallicity. To test whether a statistically significant relationship existed between \rgbf\ and [Fe/H], we split our data into separate groups that represented all GCs, those with LMXB, qLMXBs, and either an LMXB or qLMXB. Performing Spearman's rank test on each of these groups for the \rgbf\ and metallicity parameters all returned $p$-values $>0.1$, indicating we cannot reject the null hypothesis (i.e.\ there is no evidence for correlation).

\subsection{Globular Cluster Volume} \label{sec:gcvol-norm}

\begin{figure*}
\includegraphics[]{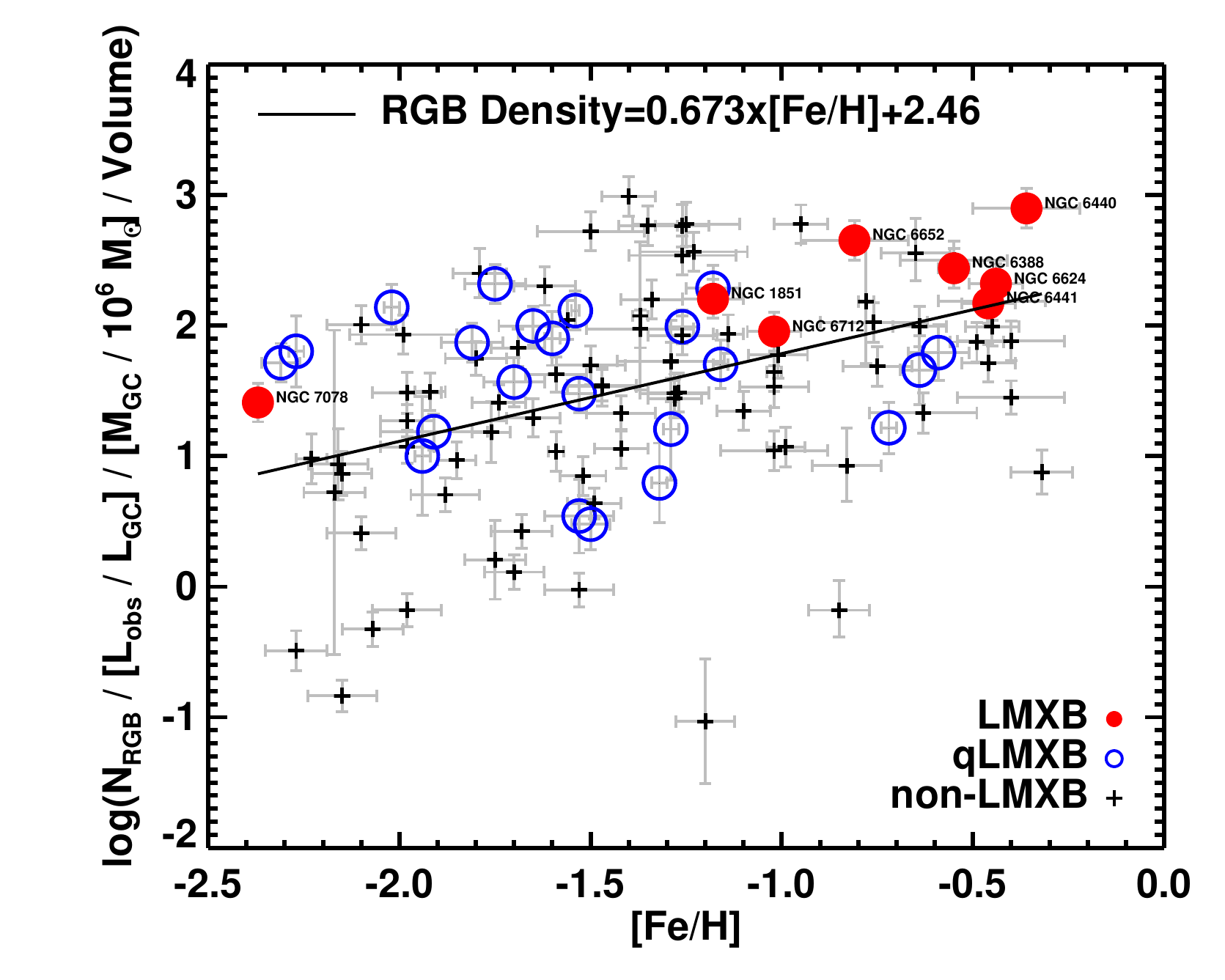}
\caption{RGB star density vs. metallicity [Fe/H]. The volume was calculated assuming spherical GCs with radius \rh. Black plusses represent normal GCs, open blue circles are GCs hosting quiescent LMXBs and filled red circles are GCs hosting bright LMXBs. NGC 6440 and NGC 7078 each have 2 LMXBs. Names of GCs with LMXBs are indicated near each datapoint. The solid line shows the weighted least squares fit (equation \ref{eq:rgbnum-massmetalvol}), which has a steeper slope than the \rgbf\ best fit. This stems from the fact that GCs with smaller $r_{h}$ are more massive. The link between \rh\ and LMXBs is evident as the metal-rich LMXB-hosting GCs (red circles) are among the GCs with the highest RGB star number density at a given [Fe/H]. The correlation is biased by the intrinsic dependence of RGB star density on $r_{h}$ (compactness), however this degeneracy is difficult to remove.}\label{fig:rgb-lfracmass-vol}
\end{figure*}

While there is no correlation between \rgbf\ and metallicity, the predictions of \citet{ivanova12-12} were based on the hypothesis that the number densities of RGB stars influence the formation rate of LMXBs. Therefore we must normalize this value by some volumetric quantity.
We used the half-light radius $r_{h}$ (in pc) to determine the volume of a cluster. Uncertainties on $r_{h}$ were taken from \citet{mclaughlin12-05} and were on average $<10\%$, where this mean value was used for GCs without uncertainties. \rh\ is a characteristic scale for the size of a GC because it is not affected by dynamical evolution in the same way $r_{c}$ is. The uncertainty associated with $r_{c}$ can be large \citep[e.g.][]{goldsbury11-13}, and the small values  for many GCs mean that our sample size of RGB stars enters the Poisson regime. $r_{c}$ is also known to be one of the strongest indicators of LMXB formation via the stellar encounter rate, and we want to exclude parameters that influence LMXB formation. One would expect the majority of RGB stars to be located near the centre of a GC due to mass segregation, as they are among the most massive members of the cluster that appear on the CMD. There are a number of caveats with using $r_{h}$ to determine the volume within which RGB stars are located. Firstly, not all RGB stars will be located within $r_{h}$. Because we only have projected distances of stars and not a 3D distribution, we can't account for the distance of stars from the GC core along the line of sight. Therefore we cannot determine what fraction of the RGB stars that we have identified will be within $r_{h}$. This would affect the number density of RGB stars since the volume within which a percentage of RGB stars resides would be different for each GC. However, while the projected and 3D distributions are not the same, this effect will average out for all clusters. We checked the radial distribution of the RGB stars we identified in each GC and most are not peaked near the core but instead can be approximated by a Gaussian. Using \rh\ means we assumed a constant density of RGB stars in the cluster to the half-light radius, which is not accurate given the radial distribution of all stars. This assumption is justified because while the stellar distribution peaks in the core, the peak of the RGB star distribution generally does not, so \rh\ as a characteristic size scale for GCs is acceptable in this case. Secondly, $r_{h}$ is another parameter, like mass, that influences the formation of LMXBs, where GCs with smaller $r_{h}$ (more compact) have been shown to preferentially host LMXBs in the Milky Way \citep{bregman03-06} and other galaxies \citep{sivakoff05-07, jordan10-04, peacock10-10, vulic08-14}. The stellar encounter rate $\Gamma$, which influences LMXB production, has a stronger correlation with LMXB occurrence when calculated using the core radius $r_{c}$ as opposed to $r_{h}$ \citep[e.g.][]{peacock10-10, bahramian04-13, agar11-13}. Even so, by using $r_{h}$ we would introduce an additional confounding effect in attempting to find a relationship between the RGB star density and [Fe/H]. This is a difficult degeneracy to remove since any measurement of the RGB density requires an estimation of the volume. In addition, $r_{h}$ has been shown to be weakly negatively correlated with [Fe/H] in M31 GCs \citep{barmby06-07}, while \citet{vanderbeke07-15} found tentative evidence for a correlation between \rh\ and [Fe/H] in Galactic GCs. \citet{vanderbeke07-15} state that this trend is caused by metal-rich GCs being more centrally concentrated than metal-poor GCs. Studies have found that most GC populations have metal-rich GCs that are on average 20\% ($\sim0.4$ pc) smaller than metal-poor GCs, likely due to different dynamical histories \citep[e.g.][]{kundu12-98, larsen06-01, jordan12-05, harris09-09, paolillo08-11}. \citet{vanderbeke07-15} concluded that the origin of the size difference (\rh) is related to the Galactocentric distance and not [Fe/H]. \citet{miocchi09-13} studied 26 Galactic GCs and also confirmed that the correlation between half-mass radius and Galactocentric radius does not depend on other cluster properties. Both studies confirm a purely dynamical origin for the correlation, suggesting tidal stripping from the bulge/disc was responsible for the correlation. If indeed \rh\ is independent of [Fe/H] and is only correlated with Galactocentric distance, then the location of a GC (and not \rh) influences its metallicity. 
Therefore, we calculated the RGB density using \rh\ and show its relationship with [Fe/H] in Figure \ref{fig:rgb-lfracmass-vol}.

The RGB star number density shows a stronger correlation with metallicity than does \nrgb.
A weighted least-squares fit to the data produced the relationship in equation \ref{eq:rgbnum-massmetalvol}. 
\begin{align}
\rm{RGB\, Density} = \log_{10}\left[\frac{N_{\rm{RGB}}}{L_{\rm{frac}}\rm{M}_{GC6}\times\frac{4}{3}\pi r_{h}^{3}}\right]		& = & \nonumber \\ (0.673 \pm 0.028)\times\rm{[Fe/H]} +  (2.46 \pm 0.04)		\label{eq:rgbnum-massmetalvol}
\end{align}
Using Spearman's rank test we found a $p$-value of 0.00016 and coefficient $r_{s} = 0.35$. Our $p$-value means we can reject the null hypothesis that the data is drawn from a random distribution, and the coefficient indicates a moderate linear relation.
We also used the Kendall rank test and found a $p$-value of 0.00021 and coefficient $\tau = 0.24$. Kendall's rank test is not as sensitive to uncertainty as Spearman's rank test but is more accurate for nonlinear correlations. 
The GCs with LMXBs preferentially have larger RGB star number densities as we expected based on their relation with $r_{h}$, and all GC-LMXBs are located above the line of best fit in Figure \ref{fig:rgb-lfracmass-vol}. The qLMXBs are more prevalent in the metal-poor population and have larger mean RGB star number densities than the rest of the metal-poor population. However, GC 47 Tucanae (NGC 104), for example, has 5 qLMXBs, and so the qLMXB distribution in our Figures does not accurately represent number statistics for individual qLMXBs but instead of the clusters within which they reside. A detailed analysis of this population is beyond the scope of this work.

\begin{figure*}
\includegraphics[]{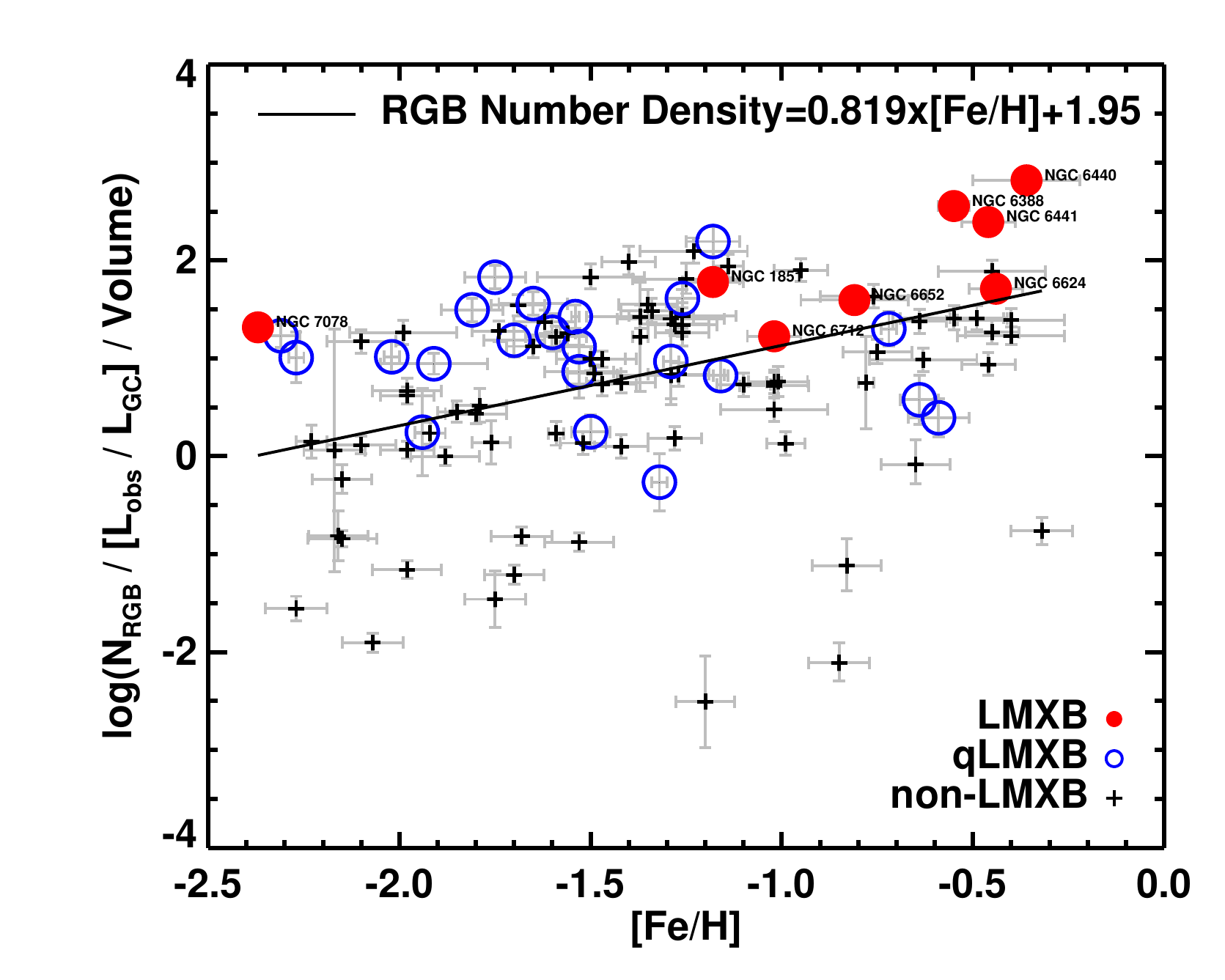}
\caption{RGB star density (not normalized by GC mass) vs. metallicity [Fe/H]. The volume was calculated using the cluster half-light radius $r_{h}$ and a spherical distribution for the GC. Black plusses represent normal GCs, open blue circles are GCs hosting quiescent LMXBs and filled red circles are GCs hosting bright LMXBs. NGC 6440 and NGC 7078 each have 2 LMXBs. Names of GCs with LMXBs are indicated near each datapoint. The solid line shows the weighted least squares fit (equation \ref{eq:rgbnum-lfracvol-metal}), which has a steeper slope than the RGB number density normalized by GC mass. As stated for Figure \ref{fig:rgb-lfracmass-vol}, $r_{h}$ affects LMXB formation and GC metallicity. The GCs with the largest RGB star density at a given metallicity are more likely to host LMXBs.}\label{fig:rgb-lfrac-vol}
\end{figure*}

In Figure \ref{fig:rgb-lfrac-vol} we plot the unnormalized RGB star density (i.e.\ RGB stars not divided by GC mass) against metallicity. This quantity best represents the number density of RGB stars as defined in \citet{ivanova12-12}. Because there is no evidence of a mass-metallicity relation for Galactic GCs, the fact that RGB stars are highly correlated with mass should not cause a metallicity effect. However, the explicit dependence of volume on $r_{h}$, which in turn affects LMXB formation and can influence metallicity, remains. In equation \ref{eq:rgbnum-lfracvol-metal} we present the weighted least squares fit from Figure \ref{fig:rgb-lfrac-vol}. A Spearman Rank test gave a $p$-value of 0.0035 and coefficient $r_{s} = 0.28$, indicating a slightly less significant correlation compared to the mass-normalized case.

\begin{align}
\rm{RGB\, Number\, Density} = \log_{10}\left[\frac{N_{\rm{RGB}}}{L_{\rm{frac}}\frac{4}{3}\pi r_{h}^{3}}\right]		& = & \nonumber \\ (0.819 \pm 0.022)\times\rm{[Fe/H]} +  (1.95 \pm 0.03)		\label{eq:rgbnum-lfracvol-metal}
\end{align}

In Table \ref{tab:rgb-values} we
report our values for the RGB fraction \rgbf\ and indicate whether the GC hosts an X-ray source. Where upper and lower limits (UL and LL) are indicated, Poisson statistics \citep{gehrels04-86} are used for values $<50$ while $\sqrt{N}$ is used for values $\ge50$.

\begin{table*}
\caption{Globular Cluster Parameters and Red Giant Branch Star Values\label{tab:rgb-values}}
\resizebox{\textwidth}{!}{%

\begin{tabular}{@{}c c  c  c  c  c  c  c  c  c  c  c  c  c  c  c  c  c  c  c  c@{}}
\hline\hline
Globular Cluster	&	$M_{\rm{V}}$	&	[Fe/H]	&	Concentration	&	$r_{c}$	&	$r_{h}$	&	$R_{\rm{Sun}}$	&	\nrgb		&	$\Delta$\nrgb (UL)	&	$\Delta$\nrgb (LL)	&	\lf	& $\Delta$\lf		&	$M_{V}$(fgstars)	&	$M/L$	&	$\Delta M/L$	&	Mass	&	$\Delta$Mass	&	\rgbf	&	$\Delta$\rgbf (UL)	&	$\Delta$\rgbf (LL)	&	X-ray Source		\\
&	mag	&	&	&	pc	&	pc	&	pc	&	&	&	&	&	&	mag	&	M$_{\odot}$/L$_{\odot}$	&	M$_{\odot}$/L$_{\odot}$	&	M$_{\odot}$	&	M$_{\odot}$	&	&	&		\\
\hline
        Arp 2 & -5.29 & -1.75 & 0.88 & 9.900 & 14.725 & 28600 & 125 & 11.180 & 11.180 & 0.270 & 0.179 & 12.56659 & 1.867 & 0.156 & 2.15E+04 & 4.34E+03 & 4.334 & 0.303 & 0.303 & -
 \\ E 3 & -4.12 & -0.83 & 0.75 & 4.406 & 4.948 & 8100 & 18 & 6.754 & 5.177 & 0.465 & 0.239 & 16.66471 & 2.303 & 0.230 & 9.02E+03 & 1.89E+03 & 3.633 & 0.291 & 0.271 & -
 \\ IC 1257 & -6.15 & -1.70 & 1.55 & 1.818 & 10.181 & 25000 & 107 & 10.344 & 10.344 & 0.394 & 0.079 & 9.50870 & 1.867 & 0.155 & 4.74E+04 & 9.58E+03 & 3.758 & 0.131 & 0.131 & -
 \\ IC 4499 & -7.32 & -1.53 & 1.21 & 4.594 & 9.351 & 18800 & 260 & 16.125 & 16.125 & 0.573 & 0.119 & 14.02200 & 1.874 & 0.154 & 1.40E+05 & 2.82E+04 & 3.511 & 0.129 & 0.129 & -
 \\ Lynga 7 & -6.60 & -1.01 & 0.95 & 2.094 & 2.548 & 7300 & 329 & 18.138 & 18.138 & 0.819 & 0.279 & 15.64936 & 2.524 & 0.276 & 9.70E+04 & 2.08E+04 & 3.617 & 0.176 & 0.176 & -
 \\ NGC 0104 & -9.42 & -0.72 & 2.07 & 0.471 & 4.150 & 4500 & 2416 & 49.153 & 49.153 & 0.403 & 0.161 & 14.01694 & 2.348 & 0.239 & 1.21E+06 & 2.55E+05 & 3.694 & 0.197 & 0.197 & qLMXB
 \\ NGC 1261 & -7.80 & -1.27 & 1.16 & 1.660 & 3.224 & 16300 & 808 & 28.425 & 28.425 & 0.838 & 0.228 & 10.10437 & 1.928 & 0.158 & 2.24E+05 & 4.51E+04 & 3.634 & 0.148 & 0.148 & -
 \\ NGC 1851 & -8.33 & -1.18 & 1.86 & 0.317 & 1.795 & 12100 & 1241 & 35.228 & 35.228 & 0.850 & 0.231 & 13.02692 & 1.981 & 0.166 & 3.75E+05 & 7.58E+04 & 3.591 & 0.148 & 0.148 & LMXB
 \\ NGC 1904 & -7.86 & -1.60 & 1.70 & 0.600 & 2.439 & 12900 & 446 & 21.119 & 21.119 & 0.398 & 0.171 & 10.74033 & 1.877 & 0.154 & 2.30E+05 & 4.64E+04 & 3.687 & 0.207 & 0.207 & qLMXB
 \\ NGC 2298 & -6.31 & -1.92 & 1.38 & 0.974 & 3.079 & 10800 & 184 & 13.565 & 13.565 & 0.876 & 0.217 & 13.16012 & 1.870 & 0.157 & 5.50E+04 & 1.11E+04 & 3.582 & 0.142 & 0.142 & -
 \\ NGC 2419 & -9.42 & -2.15 & 1.37 & 7.689 & 21.384 & 82600 & 1229 & 35.057 & 35.057 & 0.209 & 0.039 & 10.13530 & 1.903 & 0.161 & 9.82E+05 & 1.99E+05 & 3.777 & 0.121 & 0.121 & -
 \\ NGC 2808 & -9.39 & -1.14 & 1.56 & 0.698 & 2.234 & 9600 & 3308 & 57.515 & 57.515 & 0.810 & 0.220 & 12.33190 & 2.018 & 0.172 & 1.01E+06 & 2.06E+05 & 3.605 & 0.148 & 0.148 & -
 \\ NGC 288 & -6.75 & -1.32 & 0.99 & 3.495 & 5.773 & 8900 & 190 & 13.784 & 13.784 & 0.434 & 0.290 & 11.59449 & 1.972 & 0.165 & 8.70E+04 & 1.76E+04 & 3.702 & 0.305 & 0.305 & qLMXB
 \\ NGC 3201 & -7.45 & -1.59 & 1.29 & 1.853 & 4.419 & 4900 & 214 & 14.629 & 14.629 & 0.345 & 0.094 & 11.97579 & 1.876 & 0.154 & 1.58E+05 & 3.18E+04 & 3.594 & 0.150 & 0.150 & -
 \\ NGC 0362 & -8.43 & -1.26 & 1.76 & 0.450 & 2.051 & 8600 & 1060 & 32.558 & 32.558 & 0.715 & 0.152 & 8.28431 & 2.013 & 0.171 & 4.17E+05 & 8.47E+04 & 3.550 & 0.128 & 0.128 & qLMXB
 \\ NGC 4147 & -6.17 & -1.80 & 1.83 & 0.505 & 2.695 & 19300 & 209 & 14.457 & 14.457 & 0.940 & 0.200 & 10.37009 & 1.869 & 0.157 & 4.84E+04 & 9.79E+03 & 3.663 & 0.131 & 0.131 & -
 \\ NGC 4372 & -7.79 & -2.17 & 1.30 & 2.953 & 6.597 & 5800 & 76 & 8.718 & 8.718 & 0.055 & 0.157 & 11.04971 & 1.898 & 0.161 & 2.18E+05 & 4.43E+04 & 3.802 & 1.244 & 1.244 & -
 \\ NGC 4590 & -7.37 & -2.23 & 1.41 & 1.738 & 4.524 & 10300 & 253 & 15.906 & 15.906 & 0.461 & 0.177 & 12.41205 & 1.893 & 0.160 & 1.48E+05 & 3.00E+04 & 3.569 & 0.191 & 0.191 & -
 \\ NGC 4833 & -8.17 & -1.85 & 1.25 & 1.920 & 4.627 & 6600 & 483 & 21.977 & 21.977 & 0.410 & 0.099 & 9.50870 & 1.868 & 0.156 & 3.05E+05 & 6.17E+04 & 3.587 & 0.139 & 0.139 & -
 \\ NGC 5024 & -8.71 & -2.10 & 1.72 & 1.822 & 6.821 & 17900 & 1155 & 33.985 & 33.985 & 0.668 & 0.139 & 10.08142 & 1.884 & 0.159 & 5.06E+05 & 1.02E+05 & 3.534 & 0.127 & 0.127 & -
 \\ NGC 5053 & -6.76 & -2.27 & 0.74 & 10.528 & 13.210 & 17400 & 70 & 8.367 & 8.367 & 0.260 & 0.068 & 10.74033 & 1.931 & 0.164 & 8.60E+04 & 1.74E+04 & 3.496 & 0.153 & 0.153 & -
 \\ NGC 5139 & -10.26 & -1.53 & 1.31 & 3.585 & 7.563 & 5200 & 2598 & 50.971 & 50.971 & 0.197 & 0.122 & 10.13530 & 1.872 & 0.154 & 2.09E+06 & 4.23E+05 & 3.800 & 0.283 & 0.283 & qLMXB
 \\ NGC 5272 & -8.88 & -1.50 & 1.89 & 1.098 & 6.854 & 10200 & 1325 & 36.401 & 36.401 & 0.553 & 0.219 & 11.86501 & 1.877 & 0.154 & 5.89E+05 & 1.19E+05 & 3.609 & 0.194 & 0.194 & qLMXB
 \\ NGC 5286 & -8.74 & -1.69 & 1.41 & 0.953 & 2.484 & 11700 & 1951 & 44.170 & 44.170 & 0.870 & 0.221 & 10.15669 & 1.868 & 0.155 & 5.15E+05 & 1.04E+05 & 3.638 & 0.141 & 0.141 & -
 \\ NGC 5466 & -6.98 & -1.98 & 1.04 & 6.656 & 10.705 & 16000 & 128 & 11.314 & 11.314 & 0.359 & 0.069 & 11.04971 & 1.919 & 0.163 & 1.05E+05 & 2.12E+04 & 3.532 & 0.127 & 0.127 & -
 \\ NGC 5634 & -7.69 & -1.88 & 2.07 & 0.660 & 6.304 & 25200 & 434 & 20.833 & 20.833 & 0.415 & 0.086 & 12.76173 & 1.873 & 0.157 & 1.96E+05 & 3.98E+04 & 3.726 & 0.127 & 0.127 & -
 \\ NGC 5694 & -7.83 & -1.98 & 1.89 & 0.611 & 4.072 & 35000 & 634 & 25.179 & 25.179 & 0.537 & 0.101 & 10.42474 & 1.871 & 0.157 & 2.23E+05 & 4.52E+04 & 3.724 & 0.121 & 0.121 & -
 \\ NGC 5824 & -8.85 & -1.91 & 1.98 & 0.560 & 4.202 & 32100 & 1383 & 37.189 & 37.189 & 0.506 & 0.122 & 15.46308 & 1.870 & 0.157 & 5.71E+05 & 1.16E+05 & 3.680 & 0.138 & 0.138 & qLMXB
 \\ NGC 5904 & -8.81 & -1.29 & 1.73 & 0.960 & 3.862 & 7500 & 968 & 31.113 & 31.113 & 0.432 & 0.378 & 10.40342 & 1.958 & 0.163 & 5.76E+05 & 1.16E+05 & 3.590 & 0.391 & 0.391 & qLMXB
 \\ NGC 5927 & -7.81 & -0.49 & 1.60 & 0.941 & 2.464 & 7700 & 1103 & 33.211 & 33.211 & 0.684 & 0.184 & 15.77700 & 2.926 & 0.345 & 3.43E+05 & 7.50E+04 & 3.673 & 0.151 & 0.151 & -
 \\ NGC 5946 & -7.18 & -1.29 & 2.50 & 0.247 & 2.744 & 10600 & 344 & 18.547 & 18.547 & 0.590 & 0.404 & 9.61522 & 1.918 & 0.157 & 1.26E+05 & 2.54E+04 & 3.666 & 0.311 & 0.311 & -
 \\ NGC 5986 & -8.44 & -1.59 & 1.23 & 1.422 & 2.965 & 10400 & 1229 & 35.057 & 35.057 & 0.677 & 0.160 & 10.13912 & 1.876 & 0.154 & 3.93E+05 & 7.92E+04 & 3.665 & 0.136 & 0.136 & -
 \\ NGC 6093 & -8.23 & -1.75 & 1.68 & 0.436 & 1.774 & 10000 & 1286 & 35.861 & 35.861 & 0.816 & 0.222 & 10.33998 & 1.867 & 0.155 & 3.22E+05 & 6.51E+04 & 3.690 & 0.148 & 0.148 & qLMXB
 \\ NGC 6101 & -6.94 & -1.98 & 0.80 & 4.345 & 4.704 & 15400 & 278 & 16.673 & 16.673 & 0.549 & 0.106 & 10.42474 & 1.869 & 0.157 & 9.83E+04 & 1.99E+04 & 3.712 & 0.124 & 0.124 & -
 \\ NGC 6121 & -7.19 & -1.16 & 1.65 & 0.742 & 2.771 & 2200 & 140 & 11.832 & 11.832 & 0.236 & 0.087 & 9.61522 & 1.991 & 0.168 & 1.32E+05 & 2.67E+04 & 3.654 & 0.186 & 0.186 & qLMXB
 \\ NGC 6139 & -8.36 & -1.65 & 1.86 & 0.441 & 2.497 & 10100 & 904 & 30.067 & 30.067 & 0.382 & 0.104 & 8.13620 & 1.868 & 0.155 & 3.63E+05 & 7.34E+04 & 3.814 & 0.148 & 0.148 & qLMXB
 \\ NGC 6144 & -6.85 & -1.76 & 1.55 & 2.434 & 4.220 & 8900 & 232 & 15.232 & 15.232 & 0.533 & 0.265 & 9.92719 & 1.867 & 0.155 & 9.04E+04 & 1.83E+04 & 3.683 & 0.235 & 0.235 & -
 \\ NGC 6171 & -7.12 & -1.02 & 1.53 & 1.043 & 3.221 & 6400 & 323 & 17.972 & 17.972 & 0.402 & 0.109 & 13.52426 & 2.089 & 0.186 & 1.30E+05 & 2.65E+04 & 3.792 & 0.150 & 0.150 & -
 \\ NGC 6205 & -8.55 & -1.53 & 1.53 & 1.280 & 3.490 & 7100 & 1252 & 35.384 & 35.384 & 0.535 & 0.145 & 9.16737 & 1.881 & 0.154 & 4.36E+05 & 8.78E+04 & 3.730 & 0.148 & 0.148 & qLMXB
 \\ NGC 6218 & -7.31 & -1.37 & 1.34 & 1.103 & 2.471 & 4800 & 380 & 19.494 & 19.494 & 0.361 & 0.464 & 11.91284 & 1.892 & 0.155 & 1.40E+05 & 2.82E+04 & 3.876 & 0.565 & 0.565 & -
 \\ NGC 6229 & -8.06 & -1.47 & 1.50 & 1.065 & 3.194 & 30500 & 734 & 27.092 & 27.092 & 0.547 & 0.102 & 10.87180 & 1.904 & 0.156 & 2.81E+05 & 5.66E+04 & 3.680 & 0.120 & 0.120 & -
 \\ NGC 6235 & -6.29 & -1.28 & 1.53 & 1.104 & 3.345 & 11500 & 137 & 11.705 & 11.705 & 0.572 & 0.149 & 9.92719 & 1.912 & 0.157 & 5.52E+04 & 1.11E+04 & 3.637 & 0.148 & 0.148 & -
 \\ NGC 6254 & -7.48 & -1.56 & 1.38 & 0.986 & 2.496 & 4400 & 574 & 23.958 & 23.958 & 0.489 & 0.133 & 8.89801 & 1.885 & 0.154 & 1.63E+05 & 3.28E+04 & 3.857 & 0.148 & 0.148 & -
 \\ NGC 6256 & -7.15 & -1.02 & 2.50 & 0.060 & 2.577 & 10300 & 246 & 15.684 & 15.684 & 0.653 & 0.189 & 10.60710 & 2.419 & 0.254 & 1.54E+05 & 3.27E+04 & 3.388 & 0.158 & 0.158 & -
 \\ NGC 6266 & -9.18 & -1.18 & 1.71 & 0.435 & 1.820 & 6800 & 1210 & 34.785 & 34.785 & 0.307 & 0.109 & 9.58422 & 1.950 & 0.161 & 8.07E+05 & 1.63E+05 & 3.689 & 0.178 & 0.178 & qLMXB
 \\ NGC 6273 & -9.13 & -1.74 & 1.53 & 1.101 & 3.379 & 8800 & 985 & 31.385 & 31.385 & 0.321 & 0.077 & 9.22696 & 1.868 & 0.155 & 7.38E+05 & 1.49E+05 & 3.619 & 0.137 & 0.137 & -
 \\ NGC 6284 & -7.96 & -1.26 & 2.50 & 0.312 & 2.937 & 15300 & 644 & 25.377 & 25.377 & 0.276 & 0.075 & 10.89594 & 1.938 & 0.160 & 2.61E+05 & 5.26E+04 & 3.952 & 0.149 & 0.149 & -
 \\ NGC 6287 & -7.36 & -2.10 & 1.38 & 0.793 & 2.023 & 9400 & 205 & 14.318 & 14.318 & 0.396 & 0.104 & 8.75438 & 1.892 & 0.160 & 1.46E+05 & 2.97E+04 & 3.548 & 0.147 & 0.147 & -
 \\ NGC 6293 & -7.78 & -1.99 & 2.50 & 0.138 & 2.459 & 9500 & 427 & 20.664 & 20.664 & 0.374 & 0.101 & 8.91537 & 1.876 & 0.158 & 2.14E+05 & 4.33E+04 & 3.727 & 0.148 & 0.148 & -
 \\ NGC 6304 & -7.30 & -0.45 & 1.80 & 0.360 & 2.437 & 5900 & 824 & 28.705 & 28.705 & 0.733 & 0.200 & 8.39241 & 2.567 & 0.284 & 1.88E+05 & 4.04E+04 & 3.776 & 0.151 & 0.151 & -
 \\ NGC 6316 & -8.34 & -0.45 & 1.65 & 0.514 & 1.966 & 10400 & 513 & 22.650 & 22.650 & 0.208 & 0.054 & 8.89052 & 2.625 & 0.295 & 5.01E+05 & 1.08E+05 & 3.693 & 0.148 & 0.148 & -
 \\ NGC 6325 & -6.96 & -1.25 & 2.50 & 0.068 & 1.429 & 7800 & 198 & 14.071 & 14.071 & 0.251 & 0.082 & 9.82771 & 2.007 & 0.170 & 1.07E+05 & 2.18E+04 & 3.866 & 0.170 & 0.170 & -
 \\ NGC 6341 & -8.21 & -2.31 & 1.68 & 0.628 & 2.463 & 8300 & 730 & 27.019 & 27.019 & 0.686 & 0.186 & 9.95255 & 1.929 & 0.164 & 3.27E+05 & 6.63E+04 & 3.513 & 0.148 & 0.148 & qLMXB
 \\ NGC 6342 & -6.42 & -0.55 & 2.50 & 0.124 & 1.805 & 8500 & 216 & 14.697 & 14.697 & 0.343 & 0.089 & 8.63129 & 2.484 & 0.267 & 8.09E+04 & 1.73E+04 & 3.892 & 0.149 & 0.149 & -
 \\ NGC 6352 & -6.47 & -0.64 & 1.10 & 1.352 & 3.339 & 5600 & 280 & 16.733 & 16.733 & 0.475 & 0.271 & 9.07131 & 2.419 & 0.254 & 8.25E+04 & 1.75E+04 & 3.854 & 0.265 & 0.265 & qLMXB
 \\ NGC 6355 & -8.07 & -1.37 & 2.50 & 0.134 & 2.355 & 9200 & 321 & 17.916 & 17.916 & 0.220 & 0.058 & 9.85393 & 1.888 & 0.155 & 2.81E+05 & 5.67E+04 & 3.715 & 0.145 & 0.145 & -
 \\ NGC 6356 & -8.51 & -0.40 & 1.59 & 1.054 & 3.558 & 15100 & 1253 & 35.398 & 35.398 & 0.391 & 0.078 & 10.68691 & 2.703 & 0.309 & 6.03E+05 & 1.31E+05 & 3.726 & 0.128 & 0.128 & -
 \\ NGC 6362 & -6.95 & -0.99 & 1.09 & 2.498 & 4.532 & 7600 & 287 & 16.941 & 16.941 & 0.546 & 0.149 & 8.73967 & 2.160 & 0.200 & 1.15E+05 & 2.36E+04 & 3.661 & 0.150 & 0.150 & -
 \\ NGC 6366 & -5.74 & -0.59 & 0.74 & 2.209 & 2.973 & 3500 & 97 & 9.849 & 9.849 & 0.357 & 0.154 & 10.60710 & 2.282 & 0.225 & 3.97E+04 & 8.30E+03 & 3.835 & 0.213 & 0.213 & qLMXB
 \\ NGC 6380 & -7.50 & -0.75 & 1.55 & 1.078 & 2.346 & 10900 & 450 & 21.213 & 21.213 & 0.715 & 0.195 & 8.52676 & 2.703 & 0.309 & 2.38E+05 & 5.16E+04 & 3.422 & 0.153 & 0.153 & -
 \\ NGC 6388 & -9.41 & -0.55 & 1.75 & 0.346 & 1.497 & 9900 & 4003 & 63.269 & 63.269 & 0.791 & 0.215 & 9.88980 & 2.552 & 0.281 & 1.31E+06 & 2.80E+05 & 3.589 & 0.151 & 0.151 & LMXB
 \\ NGC 6397 & -6.64 & -2.02 & 2.50 & 0.033 & 1.940 & 2300 & 111 & 10.536 & 10.536 & 0.350 & 0.117 & 9.58422 & 1.879 & 0.159 & 7.49E+04 & 1.52E+04 & 3.627 & 0.175 & 0.175 & qLMXB
 \\ NGC 6401 & -7.90 & -1.02 & 1.69 & 0.771 & 5.889 & 10600 & 447 & 21.142 & 21.142 & 0.174 & 0.048 & 8.65301 & 2.135 & 0.195 & 2.72E+05 & 5.59E+04 & 3.975 & 0.150 & 0.150 & -
 \\ NGC 6402 & -9.10 & -1.28 & 0.99 & 2.137 & 3.517 & 9300 & 942 & 30.692 & 30.692 & 0.232 & 0.063 & 9.66426 & 1.915 & 0.157 & 7.36E+05 & 1.48E+05 & 3.742 & 0.148 & 0.148 & -
 \\ NGC 6426 & -6.67 & -2.15 & 1.70 & 1.558 & 5.513 & 20600 & 215 & 14.663 & 14.663 & 0.527 & 0.173 & 9.22696 & 1.925 & 0.164 & 7.89E+04 & 1.60E+04 & 3.714 & 0.170 & 0.170 & -
 \\ NGC 6440 & -8.75 & -0.36 & 1.62 & 0.346 & 1.187 & 8500 & 1399 & 37.403 & 37.403 & 0.303 & 0.083 & 9.87176 & 2.981 & 0.354 & 8.30E+05 & 1.82E+05 & 3.745 & 0.152 & 0.152 & LMXB
 \\ NGC 6441 & -9.63 & -0.46 & 1.74 & 0.439 & 1.923 & 11600 & 5777 & 76.007 & 76.007 & 0.785 & 0.214 & 10.09970 & 2.656 & 0.301 & 1.66E+06 & 3.60E+05 & 3.646 & 0.151 & 0.151 & LMXB
 \\ NGC 6453 & -7.22 & -1.50 & 2.50 & 0.169 & 1.485 & 11600 & 719 & 26.814 & 26.814 & 0.775 & 0.211 & 10.00285 & 1.883 & 0.154 & 1.28E+05 & 2.58E+04 & 3.860 & 0.148 & 0.148 & -
 \\ NGC 6496 & -7.20 & -0.46 & 0.70 & 3.123 & 3.353 & 11300 & 223 & 14.933 & 14.933 & 0.164 & 0.039 & 8.75438 & 2.497 & 0.270 & 1.67E+05 & 3.56E+04 & 3.912 & 0.142 & 0.142 & -
 \\ NGC 6517 & -8.25 & -1.23 & 1.82 & 0.185 & 1.542 & 10600 & 590 & 24.290 & 24.290 & 0.309 & 0.084 & 8.06636 & 1.921 & 0.158 & 3.38E+05 & 6.81E+04 & 3.753 & 0.148 & 0.148 & -
 \\ NGC 6522 & -7.65 & -1.34 & 2.50 & 0.112 & 2.240 & 7700 & 569 & 23.854 & 23.854 & 0.395 & 0.108 & 12.37469 & 1.901 & 0.156 & 1.92E+05 & 3.88E+04 & 3.874 & 0.148 & 0.148 & -
 \\ NGC 6535 & -4.75 & -1.79 & 1.33 & 0.712 & 1.681 & 6800 & 50 & 10.340 & 8.822 & 0.764 & 0.254 & 8.91537 & 1.868 & 0.156 & 1.31E+04 & 2.64E+03 & 3.700 & 0.191 & 0.186 & -
 \\ NGC 6539 & -8.29 & -0.63 & 1.74 & 0.862 & 3.857 & 7800 & 296 & 17.205 & 17.205 & 0.127 & 0.035 & 16.01500 & 2.470 & 0.265 & 4.50E+05 & 9.60E+04 & 3.713 & 0.152 & 0.152 & -
 \\ NGC 6540 & -6.35 & -1.35 & 2.50 & 0.046 & 1.156 & 5300 & 176 & 13.266 & 13.266 & 0.765 & 0.208 & 13.30663 & 1.991 & 0.168 & 6.08E+04 & 1.23E+04 & 3.578 & 0.151 & 0.151 & -
 \\ NGC 6541 & -8.52 & -1.81 & 1.86 & 0.393 & 2.313 & 7500 & 832 & 28.844 & 28.844 & 0.513 & 0.139 & 10.25138 & 1.869 & 0.157 & 4.21E+05 & 8.53E+04 & 3.586 & 0.148 & 0.148 & qLMXB
 \\ NGC 6544 & -6.94 & -1.40 & 1.63 & 0.044 & 1.056 & 3000 & 90 & 9.487 & 9.487 & 0.189 & 0.051 & 10.29180 & 1.878 & 0.154 & 9.87E+04 & 1.99E+04 & 3.684 & 0.154 & 0.154 & -
 \\ NGC 6569 & -8.28 & -0.76 & 1.31 & 1.110 & 2.537 & 10900 & 664 & 25.768 & 25.768 & 0.226 & 0.062 & 10.18867 & 2.242 & 0.217 & 4.05E+05 & 8.43E+04 & 3.861 & 0.151 & 0.151 & -
 \\ NGC 6584 & -7.69 & -1.50 & 1.47 & 1.021 & 2.867 & 13500 & 486 & 22.045 & 22.045 & 0.499 & 0.136 & 11.44562 & 1.890 & 0.155 & 1.98E+05 & 4.00E+04 & 3.692 & 0.148 & 0.148 & -
 \\ NGC 6624 & -7.49 & -0.44 & 2.50 & 0.138 & 1.884 & 7900 & 892 & 29.866 & 29.866 & 0.619 & 0.169 & 9.60132 & 2.802 & 0.326 & 2.44E+05 & 5.33E+04 & 3.770 & 0.152 & 0.152 & LMXB
 \\ NGC 6637 & -7.64 & -0.64 & 1.38 & 0.845 & 2.150 & 8800 & 793 & 28.160 & 28.160 & 0.794 & 0.224 & 11.84030 & 2.419 & 0.254 & 2.42E+05 & 5.14E+04 & 3.615 & 0.154 & 0.154 & -
 \\ NGC 6638 & -7.12 & -0.95 & 1.33 & 0.602 & 1.395 & 9400 & 450 & 21.213 & 21.213 & 0.498 & 0.131 & 10.56374 & 2.127 & 0.193 & 1.32E+05 & 2.71E+04 & 3.835 & 0.146 & 0.146 & -
 \\ NGC 6642 & -6.66 & -1.26 & 1.99 & 0.236 & 1.720 & 8100 & 184 & 13.565 & 13.565 & 0.319 & 0.087 & 10.53964 & 1.928 & 0.158 & 7.83E+04 & 1.58E+04 & 3.867 & 0.151 & 0.151 & -
 \\ NGC 6652 & -6.66 & -0.81 & 1.80 & 0.291 & 1.396 & 10000 & 316 & 17.776 & 17.776 & 0.703 & 0.191 & 12.95863 & 2.151 & 0.198 & 8.74E+04 & 1.80E+04 & 3.711 & 0.150 & 0.150 & LMXB
 \\ NGC 6656 & -8.50 & -1.70 & 1.38 & 1.238 & 3.128 & 3200 & 671 & 25.904 & 25.904 & 0.342 & 0.132 & 8.89052 & 1.870 & 0.154 & 4.14E+05 & 8.35E+04 & 3.676 & 0.189 & 0.189 & qLMXB
 \\ NGC 6681 & -7.12 & -1.62 & 2.50 & 0.079 & 1.859 & 9000 & 448 & 21.166 & 21.166 & 0.705 & 0.192 & 12.15687 & 1.886 & 0.154 & 1.17E+05 & 2.36E+04 & 3.735 & 0.148 & 0.148 & -
 \\ NGC 6712 & -7.50 & -1.02 & 1.05 & 1.525 & 2.669 & 6900 & 296 & 17.205 & 17.205 & 0.221 & 0.060 & 12.46119 & 2.111 & 0.190 & 1.86E+05 & 3.81E+04 & 3.857 & 0.150 & 0.150 & LMXB
 \\ NGC 6715 & -9.98 & -1.49 & 2.04 & 0.694 & 6.321 & 26500 & 5578 & 74.686 & 74.686 & 0.747 & 0.139 & 9.82771 & 1.876 & 0.154 & 1.62E+06 & 3.27E+05 & 3.663 & 0.119 & 0.119 & -
 \\ NGC 6717 & -5.66 & -1.26 & 2.07 & 0.165 & 1.404 & 7100 & 108 & 10.392 & 10.392 & 0.506 & 0.152 & 8.63129 & 1.950 & 0.161 & 3.15E+04 & 6.37E+03 & 3.830 & 0.163 & 0.163 & -
 \\ NGC 6723 & -7.83 & -1.10 & 1.11 & 2.101 & 3.872 & 8700 & 686 & 26.192 & 26.192 & 0.522 & 0.142 & 11.10416 & 2.036 & 0.176 & 2.43E+05 & 4.94E+04 & 3.733 & 0.148 & 0.148 & -
 \\ NGC 6752 & -7.73 & -1.54 & 2.50 & 0.198 & 2.222 & 4000 & 526 & 22.935 & 22.935 & 0.429 & 0.117 & 11.63855 & 1.878 & 0.154 & 2.04E+05 & 4.12E+04 & 3.779 & 0.148 & 0.148 & qLMXB
 \\ NGC 6760 & -7.84 & -0.40 & 1.65 & 0.732 & 2.734 & 7400 & 558 & 23.622 & 23.622 & 0.265 & 0.072 & 11.63855 & 2.671 & 0.304 & 3.22E+05 & 6.97E+04 & 3.816 & 0.152 & 0.152 & -
 \\ NGC 6779 & -7.41 & -1.98 & 1.38 & 1.203 & 3.008 & 9400 & 422 & 20.543 & 20.543 & 0.794 & 0.238 & 9.85393 & 1.878 & 0.158 & 1.52E+05 & 3.08E+04 & 3.543 & 0.158 & 0.158 & -
 \\ NGC 6809 & -7.57 & -1.94 & 0.93 & 2.827 & 4.445 & 5400 & 216 & 14.697 & 14.697 & 0.333 & 0.341 & 8.52676 & 1.868 & 0.156 & 1.75E+05 & 3.55E+04 & 3.568 & 0.454 & 0.454 & qLMXB
 \\ NGC 6838 & -5.61 & -0.78 & 1.15 & 0.733 & 1.943 & 4000 & 135 & 11.619 & 11.619 & 0.778 & 0.837 & 8.65301 & 2.383 & 0.247 & 3.68E+04 & 7.78E+03 & 3.674 & 0.478 & 0.478 & -
 \\ NGC 6864 & -8.57 & -1.29 & 1.80 & 0.547 & 2.797 & 20900 & 1051 & 32.419 & 32.419 & 0.453 & 0.123 & 11.63855 & 2.013 & 0.171 & 4.75E+05 & 9.63E+04 & 3.689 & 0.148 & 0.148 & -
 \\ NGC 6934 & -7.45 & -1.47 & 1.53 & 0.998 & 3.131 & 15600 & 539 & 23.216 & 23.216 & 0.772 & 0.211 & 11.63855 & 1.881 & 0.154 & 1.58E+05 & 3.19E+04 & 3.645 & 0.149 & 0.149 & -
 \\ NGC 6981 & -7.04 & -1.42 & 1.21 & 2.275 & 4.599 & 17000 & 405 & 20.125 & 20.125 & 0.789 & 0.214 & 11.63855 & 1.912 & 0.157 & 1.10E+05 & 2.22E+04 & 3.668 & 0.149 & 0.149 & -
 \\ NGC 7006 & -7.67 & -1.52 & 1.41 & 2.037 & 5.273 & 41200 & 761 & 27.586 & 27.586 & 0.909 & 0.248 & 11.63855 & 1.871 & 0.154 & 1.93E+05 & 3.89E+04 & 3.638 & 0.148 & 0.148 & -
 \\ NGC 7078 & -9.19 & -2.37 & 2.29 & 0.424 & 3.025 & 10400 & 1403 & 37.457 & 37.457 & 0.583 & 0.159 & 11.63855 & 1.925 & 0.164 & 8.04E+05 & 1.63E+05 & 3.476 & 0.148 & 0.148 & LMXB
 \\ NGC 7089 & -9.03 & -1.65 & 1.59 & 1.070 & 3.546 & 11500 & 1855 & 43.070 & 43.070 & 0.749 & 0.204 & 11.63855 & 1.872 & 0.154 & 6.75E+05 & 1.36E+05 & 3.565 & 0.147 & 0.147 & -
 \\ NGC 7099 & -7.45 & -2.27 & 2.50 & 0.141 & 2.427 & 8100 & 332 & 18.221 & 18.221 & 0.544 & 0.322 & 9.87176 & 1.903 & 0.161 & 1.60E+05 & 3.24E+04 & 3.581 & 0.273 & 0.273 & qLMXB
 \\ Palomar 1 & -2.52 & -0.65 & 2.57 & 0.032 & 1.485 & 11100 & 11 & 5.594 & 3.978 & 0.970 & 0.254 & 10.00285 & 2.552 & 0.281 & 2.29E+03 & 4.91E+02 & 3.695 & 0.265 & 0.215 & -
 \\ Palomar 12 & -4.47 & -0.85 & 2.98 & 0.111 & 9.506 & 19000 & 21 & 7.183 & 5.617 & 0.753 & 0.250 & 12.56659 & 2.168 & 0.201 & 1.17E+04 & 2.42E+03 & 3.377 & 0.226 & 0.206 & -
 \\ Palomar 15 & -5.51 & -2.07 & 0.60 & 15.743 & 14.431 & 45100 & 128 & 11.314 & 11.314 & 0.813 & 0.168 & 16.66471 & 1.874 & 0.158 & 2.64E+04 & 5.35E+03 & 3.776 & 0.131 & 0.131 & -
 \\ Palomar 2 & -7.97 & -1.42 & 1.53 & 1.345 & 3.956 & 27200 & 975 & 31.225 & 31.225 & 0.665 & 0.154 & 14.02200 & 1.946 & 0.161 & 2.64E+05 & 5.34E+04 & 3.744 & 0.134 & 0.134 & -
 \\ Pyxis & -5.73 & -1.20 & 1.60 & 14.899 & 22.349 & 39400 & 83 & 9.110 & 9.110 & 0.569 & 0.611 & 11.59449 & 1.946 & 0.161 & 3.36E+04 & 6.78E+03 & 3.638 & 0.477 & 0.477 & -
 \\ Ruprecht 106 & -6.35 & -1.68 & 0.70 & 6.167 & 6.475 & 21200 & 126 & 11.225 & 11.225 & 0.730 & 0.143 & 13.16012 & 1.868 & 0.155 & 5.70E+04 & 1.15E+04 & 3.481 & 0.128 & 0.128 & -
 \\ Terzan 7 & -5.01 & -0.32 & 0.93 & 3.250 & 5.107 & 22800 & 80 & 8.944 & 8.944 & 0.829 & 0.249 & 10.37009 & 2.581 & 0.287 & 2.29E+04 & 4.94E+03 & 3.624 & 0.167 & 0.167 & -
 \\ Terzan 8 & -5.07 & -2.16 & 0.60 & 7.650 & 7.268 & 26300 & 147 & 12.124 & 12.124 & 0.594 & 0.347 & 12.41205 & 1.885 & 0.159 & 1.77E+04 & 3.59E+03 & 4.145 & 0.271 & 0.271 & -
 \\ 
\hline
\end{tabular}
}
\begin{list}{}{}
\item Globular cluster absolute magnitude \mv, metallicity [Fe/H], concentration, core radius $r_{\rm{c}}$, half-light radius $r_{\rm{h}}$, and distance $R_{\rm{Sun}}$ are all taken from \citetalias{harris10-96}. \nrgb\ is the number of RGB stars along with corresponding upper and lower limits (UL and LL), which are derived using $\sqrt{N}$ for values $\ge50$ and Poisson statistics \citep{gehrels04-86} for values $<50$. \lf\ is the fraction of the total cluster luminosity that was observed in the field of view of the given \HSTt\ instrument. $M_{V}$(fgstars) is the absolute $V$ magnitude from foreground stars in the cluster field of view obtained from the Besan\c{c}on models \citep{robin10-03}. Mass-to-light ratios are taken from Table 8 (column 6) of \citet{mclaughlin12-05}. Masses were derived by multiplying the $M/L$ by total cluster luminosity derived from \mv. $RGB_{\rm{frac}} =$ \nrgb/\rm{Mass}/\lf\ is the normalized number of RGB stars in a cluster.  	

\end{list}
\end{table*}

\section{Impact of RGB Stars on Globular Cluster LMXBs} \label{sec:gclmxb-rgb}

Our sample contains 10 LMXBs in 8 different GCs and 22 GCs with at least one qLMXB. Figure \ref{fig:rgb-lfrac-vol} shows that the number density of RGB stars is correlated with the metallicity of a GC. This confirms the prediction of \citet{ivanova12-12} that the number density of RGB stars is larger in metal-rich GCs, and therefore a key contributor to the dynamical formation of LMXBs. However, the correlations and degeneracies between GC parameters and LMXB formation means that RGB star number density is not the only cause of the metallicity effect. To assess the impact of RGB star density on LMXBs we need to identify their properties, such as confirmed optical counterparts and \lx.

\subsection{qLMXB Contribution}

Unlike LMXBs, qLMXBs are not expected to have red giants as donors (or seeds of formation) because they are not bright persistent sources. While qLMXBs have \lx\ $\lesssim10^{35}$ \es, XRB duty cycles in general are not well-known and thus qLMXBs could be in a transition state. Based on the low X-ray luminosities and hence low mass transfer rates, they should have main sequence or white dwarf companions (not ultracompact, i.e.\ their orbital periods are $>1$ h). Searches for optical counterparts to qLMXBs have generally only proposed candidates \citep[e.g.][]{heinke06-05, heinke02-09, maxwell09-12}. This does not rule out the possibility that some have RGB star companions, although this scenario would require a large orbital separation and/or reduced mass loss rates. However, no relationship between qLMXBs and mass nor metallicity in the Galaxy (extragalactic observations are not sensitive enough to detect qLMXBs) has been found \citep{heinke11-03}. Therefore a relationship between qLMXBs and RGB star density is not expected to exist, and as Figures \ref{fig:rgb-lfracmass} and \ref{fig:rgb-lfracmass-vol} show the distribution of qLMXB clusters is not metallicity dependent. A KS test of the \rgbf\ parameter for qLMXBs and non-LMXBs yielded a $p$-value of 0.66 and for the RGB star density parameter a $p$-value of 0.68, meaning we cannot reject the null hypothesis of uncorrelated values for either case. This result requires more investigation to analyse the qLMXB frequency in each GC and metallicity. The qLMXBs with the largest RGB star numbers (NGC 5139, NGC 6139, NGC 6352, NGC 6366) and densities (NGC 6093, NGC 6266, NGC 6388) in Figures \ref{fig:rgb-lfracmass} and \ref{fig:rgb-lfracmass-vol} are possible transient sources that have previously been in outburst, and thus good candidates for monitoring. 

We also analysed the distribution of qLMXBs to study their dependence on metallicity. When we split the qLMXBs into populations based on metallicity at [Fe/H] $= -1.0$ (the approximate separation of the bimodal [Fe/H] distribution; \citealt{bellazzini02-95}), we found the fraction of GC-qLMXBs was 14\% larger (as a fraction of the total GCs in that population) in the metal-poor (31\%) vs.\ the metal-rich (17\%) population. The number of qLMXBs within a GC can vary (see Section \ref{sec:gcvol-norm}), and so this result is only based on the number of GCs that host at least one qLMXB. In addition, selection effects are a large source of uncertainty because many GCs have not been observed at the \lx\ limits for detecting qLMXBs. We have not included statistics for multiple qLMXBs within one GC. More work into the nature of qLMXBs and GC metallicity is needed. As \citet{ivanova05-08} states, an independent correlation (or lack thereof) between qLMXB number and metallicity can help us better understand the metallicity dependence on LMXB formation.

\subsection{GC-LMXB Metallicity Relation}

While qLMXBs are not known to be dependent on GC metalliticity, LMXB formation is. However, metallicity is known to be related to many other GC parameters, such as mass and density. Despite this, the metallicity effect has been shown to be independent of other factors. \citet{kim02-13} showed that for a sample of 408 extragalactic GC-LMXBs, the mass-metallicity relation was negligible (15\%) compared to the factor of 3 difference in LMXB production between metal-rich and metal-poor GCs. 
\citet{kim02-13} also found that for extragalactic LMXBs, the metallicity dependence not only hold for all bright LMXBs, but that the effect is independent of X-ray luminosity ($>10^{36}$ \es), stellar age, dynamical properties (e.g.\ stellar encounter rate, Galactocentric distance), and selection effects. 
This makes the prediction that RGB star density causes the LMXB-metallicity dependence more intriguing, since the remaining GC parameters have been separated from the metallicity dependence (there is no known mass-metallicity relationship for Milky Way GCs). The results of our study are only physically relevant if the initial conjecture is true: that LMXB companions should be RGB stars.
Therefore we need to identify what fraction, if any, of the LMXBs in our sample are actually known to have RGB star counterparts.

\subsection{Milky Way GC-LMXB Counterparts}

An issue with work of this nature is that it is incredibly difficult to definitively determine the companions of LMXBs in GCs because the stellar density is so high. \citet{verbunt04-06} summarized potential candidates for a number of GC-LMXBs in the literature, with most being classified as a `faint star' or having no clear counterpart. The LMXB in NGC 6624 could be a white dwarf or stripped core of an evolved main sequence star. NGC 7078 (M15) has 2 LMXBs, where X-1 is thought to have a red giant companion and X-2 a blue star as its companion.
The orbital period of an LMXB was often used as an indicator of the type of companion, with orbital periods $<1$ h meaning degenerate stars, 3 h $\le P_{orb} \le 10$ h indicating main sequence stars, and $>10$ h for giants. While these studies have suggested a number of LMXB companions are main sequence stars, this is based on the identification of a blue optical counterpart near/at the position of the X-ray source. The accretion disc of an LMXB is known to emit in the UV, which further complicates optical counterpart identification.
With no confirmation of counterparts for any LMXBs, we cannot say with any certainty that RGB stars are or are not companions. Population synthesis models \cite[e.g.][]{ivanova05-08} have predicted that the LMXB formation rate is highest for neutron stars with main sequence donors. However, the mass transfer rates and thus X-ray luminosities of these systems are low ($<10^{37}$ \es). Therefore red giant companions are thought to compose the brightest LMXBs because they provide the necessary mass transfer rates to drive higher \lx\ \citep{fragos08-08, fragos09-09}. Only three of the bright GC-LMXBs in the Milky Way have X-ray luminosities $\gtrsim10^{37}$ \es, the sources in NGC 6441, NGC 6624, and Liller 1 \citep{liu06-07}, where Liller 1 is not part of our sample.
From these candidates and their \lx\ values we can infer that 2 of the 10 LMXBs in different clusters from our sample possibly have RGB star companions.

\subsection{LMXB Compact Object Type}

We still have not addressed the impact that the accretor in the LMXB has on our interpretation. If the compact object in a GC-LMXB is a black hole, those with luminosities $>10^{37}$ \es\ don't require RGB star companions. In any case, the compact objects in bright Milky Way GC-LMXBs are all neutron stars. Only 4 black hole candidates have been identified to date, with two detected in radio and not X-ray \citep{strader10-12}, and two with X-ray luminosities $<10^{33}$ \es\ \citep{chomiuk11-13, miller-jones11-15}.
Nonetheless, \citet{ivanova07-10} found that RGB stars have a similar effect on the production of black hole LMXBs indirectly by increasing the formation rates of the seeds (LMXBs with red giant donors) of black hole-white dwarf binaries.

\subsection{RGB Star Proxies and Metallicity}

Our study attempted to address why metal-rich GCs produce more LMXBs than metal-poor GCs by investigating the relationship between the number density of RGB stars in a GC and GC metallicity. Even if our sample of GC-LMXBs does not have RGB star donors, we are still probing the relationship between the number density of RGB stars and metallicity of all GCs and not just those with an LMXB. If all metal-rich GCs had much higher RGB star densities than metal-poor ones, it would be evidence to support the impact of GC metallicity on LMXB formation (we point out that \citetalias{nataf04-13} did find a trend of increasing number counts on the RGB bump with increasing metallicity).
In Figure \ref{fig:rgb-lfracmass-vol}, the RGB star number density does appear to be larger for metal-rich LMXBs compared to the best-fitting relation, and the overall distribution is correlated with [Fe/H] based on Spearman and Kendall Rank tests. However, we caution that the explicit dependence of volume on $r_{h}$, which is correlated with both LMXB formation and metallicity, biases this result. As we stated above, the correlation of \rh\ with [Fe/H] is purely dynamical as a result of Galactocentric distance effects from tidal interactions.
When investigating the effect of RGB star density on GC metallicity, it is difficult to remove the correlation that density (through $r_{h}$) has on metallicity.

To exclude the intrinsic dependence of density on metallicity, we can use the \rgbf\ parameter. While \rgbf\ only represents the number of RGB stars per \msun\ and not a volume density, it is independent of mass and density, which both influence LMXB formation. Therefore \rgbf\ can be used as an independent probe of whether the number of RGB stars in a GC varies with metallicity, explaining the enhanced production of LMXBs in metal-rich GCs. From Figure \ref{fig:rgb-lfracmass}, we found no dependence between the \rgbf\ parameter and metallicity. Although we cannot reject the null hypothesis based on a Spearman's Rank test, this does not mean that \rgbf\ and [Fe/H] are not correlated. The shallow slope from the least squares fit and large uncertainties on \rgbf\ are an indication that further analysis with improved measurements would lead to a more robust result. As a result we cannot claim there is an independent relationship between RGB star number density and metallicity.

\subsection{Degeneracy of GC Parameters Affecting LMXB Formation}

Our goal was to determine the underlying physical cause for the metallicity dependence of GC-LMXBs. Both the number of RGB stars and the density of RGB stars are related to the GC mass (Figure \ref{fig:nrgb-gcmass}) and \rh\ (Section \ref{sec:gcvol-norm}) respectively. These parameters (GC mass and \rh) are already known to be indicators of whether a GC hosts an LMXB. Therefore the argument invokes parameters that are known to affect LMXB formation to explain the metallicity dependence.
Given that there is no mass-metallicity relation in the Galaxy, we have removed GC mass from our correlation by normalizing for it. However, the dependence of \rh\ on metallicity is confounding because it is instead attributed to a dependence of \rh\ on Galactocentric distance. The purely dynamical origin for this relationship (i.e.\ tidal interactions/stripping) is telling since dynamical interactions promote LMXB formation. If metal-rich GCs are more likely to have an active dynamical history, possibly due to a highly inclined orbit about the disc in a galaxy or the surrounding environment, they will preferentially form LMXBs. \citet{kim08-06} showed that GCs closer to the centres of galaxies are more likely to harbour LMXBs than those in the outskirts, due to Galactic tidal forces that cause GCs to have smaller core radii (higher central densities). However, a study of NGC 1399 by \citet{paolillo08-11} found that GC-LMXBs follow the radial distribution of their parent GC population and argued against any external dynamical effects from the galaxy influencing LMXB formation in GCs. Tidal stripping from GCs usually occurs in galaxy centres and has not been confirmed at larger Galactocentric distances, while the metallicity effect for LMXBs holds at all Galactocentric distances \citep{kim02-13, mineo01-14}. If external dynamics do contribute to the GC-LMXB metallicity effect, the dynamical formation/evolution of metal-rich and metal-poor GCs would have to be different throughout a galaxy. 
Therefore, while we confirmed the prediction from \citet{ivanova12-12} that the number density of RGB stars is larger in metal-rich GCs, we cannot claim that it is the (sole) underlying cause of the metallicity effect on LMXB formation.

\section{Improvements and Future Work} \label{sec:gcrgb-improvefw}

To improve our analysis a number of developments need to take place. Firstly, the uncertainties on measurements of GC distances and luminosities need to be improved. The end-of-mission $\emph{GAIA}$ data will improve distance measurements for GCs, which is currently one of the dominant sources of uncertainty in all GC studies. Higher spatial resolution data using $\emph{JWST}$ that covers the entire GC population will improve both number statistics and consistency between GCs. Specifically, observing the complete extent of each GC will reduce the large uncertainties associated with luminosity correction.
The $M/L$ ratios we used can be inaccurate but are internally consistent, only resulting in a systematic error in GC mass. However, our largest source of uncertainty in \rgbf\ comes from the modelled $M/L$ ratios used to determine masses, which limits the both the interpretation and significance of our results. Follow-up work to localize LMXBs with \Chandra\ and analyse the optical counterparts in \HSTt\ images in the X-ray error ellipse can yield further insight on companions.

Future endeavours will need to reduce the uncertainty on GC parameters in order to obtain more robust results. The GCs in M31 might appear to be ideal targets for many reasons. First, each GC is at the same distance and data is available for the entire cluster, eliminating two of the largest uncertainties in our work. The recently completed Panchromatic $\emph{Hubble}$ Andromeda Treasury survey obtained 6-filter photometry in the UV to NIR for one-third of M31's disc, and is complete for the bright-end of the RGB in the field. However, the central stellar density in M31's $\approx500$ GCs is too high to count RGB stars with current observations. Therefore our analysis remains restricted to our Galaxy and improving measurements such as $M/L$. As \citet{ivanova12-12} stated, a population synthesis study that includes red giants will also advance our understanding of their effect while controlling for GC parameters such as the affect $r_{h}$ has on RGB star density. This could be accomplished using models with the same $r_{h}$ over a wide range of metallicity at fixed mass and Galactocentric distance. The second part of the prediction from \citet{ivanova12-12} stated that the average masses of RGB stars should be larger in metal-rich GCs. This could be investigated observationally using estimates of stellar masses from isochrone-fitting across the red giant branch or spectral analysis. 
Lastly, with the upcoming launch of $\emph{eROSITA}$, which will detect all XRBs in our galaxy down to $\sim10^{33}$ \es, it will be possible to survey qLMXBs in GCs and determine any relationship between their formation or luminosity distribution and GC parameters such as metallicity.

\section{Summary}	\label{sec:summary}

One of the unanswered questions in GC-LMXB studies is the origin of the metallicity effect. Why are there $\sim3$ times more LMXBs in metal-rich GCs compared to metal-poor ones? In this work we investigated this relationship using the hypothesis that the RGB star number density is correlated with metallicity, and thus is the underlying physical cause of the correlation. We used \HSTt\ data from the ACS and WFPC2 instruments for 109 unique Milky Way GCs to calculate the number of RGB stars, \nrgb. We made corrections for the fraction of cluster light observed and foreground star contamination. Normalizing by GC mass we found no correlation between \rgbf\ (number of RGB stars per \msun) and metallicity. Because RGB stars are likely mass-segregated, many will be located within the half-light radius of a GC. We normalized \rgbf\ by the GC volume at the half-light radius to find the number density of RGB stars. The RGB star number density was correlated with metallicity [Fe/H], indicating the underlying cause of the LMXB preference for metal-rich GCs. Spearman and Kendall Rank tests gave $p$-values of 0.00016 and 0.00021 and coefficients $r_{s} = 0.35$ and $\tau = 0.24$ respectively.

However, we caution that this result is inherently biased by the half-light radius $r_{h}$, which affects LMXB formation rate and is possibly negatively correlated with GC metallicity. The dynamical origin of the \rh-metallicity correlation (tidal stripping) suggests that metal-rich GCs may have had more active dynamical histories, which would promote LMXB formation. In addition, not all LMXBs have RGB star companions, but this does not preclude a relationship between RGB star density and [Fe/H]. No correlation between qLMXBs (number or RGB star parameters) and GC metallicity was found, although a qLMXB census in Galactic GCs is needed to further this analysis. Follow-up observations of Milky Way GCs with $\emph{JWST}$ that cover the entire cluster extent will reduce uncertainties, as will updated distance measurements with $\emph{GAIA}$. Even next generation space-based optical telescopes will not have the capability to study promising extragalactic GCs in e.g.\ M31, where a more consistent, relevant, and robust analysis would be possible. An investigation of the average masses of RGB stars in relation to Milky Way GC metallicity would provide further insight into this intriguing problem.

\section*{acknowledgements}
We thank the referee for helpful comments that improved the manuscript. We thank David M. Nataf for helpful comments on the methodology and Bill Harris for comments on the manuscript. Support for this work was provided by Discovery Grants from the Natural Sciences and Engineering Research Council of Canada and by Ontario Early Researcher Awards. NV acknowledges support from Ontario Graduate Scholarships. This work was made possible by the facilities of the Shared Hierarchical Academic Research Computing Network (SHARCNET:www.sharcnet.ca) and Compute/Calcul Canada. We acknowledge the following archives: the Hubble Legacy Archive (\url{hla.stsci.edu}). \\ 
\indent \emph{Facilities:} HST (ACS, WFC, WFPC2)

\bibliographystyle{mnras}

\begin{thebibliography}{}
\makeatletter
\relax
\def\mn@urlcharsother{\let\do\@makeother \do\$\do\&\do\#\do\^\do\_\do\%\do\~}
\def\mn@doi{\begingroup\mn@urlcharsother \@ifnextchar [ {\mn@doi@}
  {\mn@doi@[]}}
\def\mn@doi@[#1]#2{\def\@tempa{#1}\ifx\@tempa\@empty \href
  {http://dx.doi.org/#2} {doi:#2}\else \href {http://dx.doi.org/#2} {#1}\fi
  \endgroup}
\def\mn@eprint#1#2{\mn@eprint@#1:#2::\@nil}
\def\mn@eprint@arXiv#1{\href {http://arxiv.org/abs/#1} {{\tt arXiv:#1}}}
\def\mn@eprint@dblp#1{\href {http://dblp.uni-trier.de/rec/bibtex/#1.xml}
  {dblp:#1}}
\def\mn@eprint@#1:#2:#3:#4\@nil{\def\@tempa {#1}\def\@tempb {#2}\def\@tempc
  {#3}\ifx \@tempc \@empty \let \@tempc \@tempb \let \@tempb \@tempa \fi \ifx
  \@tempb \@empty \def\@tempb {arXiv}\fi \@ifundefined
  {mn@eprint@\@tempb}{\@tempb:\@tempc}{\expandafter \expandafter \csname
  mn@eprint@\@tempb\endcsname \expandafter{\@tempc}}}

\bibitem[\protect\citeauthoryear{{Agar} \& {Barmby}}{{Agar} \&
  {Barmby}}{2013}]{agar11-13}
{Agar} J.~R.~R.,  {Barmby} P.,  2013, \mn@doi [\aj]
  {10.1088/0004-6256/146/5/135}, \href
  {http://adsabs.harvard.edu/abs/2013AJ....146..135A} {146, 135}

\bibitem[\protect\citeauthoryear{{Bahramian}, {Heinke}, {Sivakoff}  \&
  {Gladstone}}{{Bahramian} et~al.}{2013}]{bahramian04-13}
{Bahramian} A.,  {Heinke} C.~O.,  {Sivakoff} G.~R.,   {Gladstone} J.~C.,  2013,
  \mn@doi [\apj] {10.1088/0004-637X/766/2/136}, \href
  {http://adsabs.harvard.edu/abs/2013ApJ...766..136B} {766, 136}

\bibitem[\protect\citeauthoryear{{Bahramian} et~al.,}{{Bahramian}
  et~al.}{2014}]{bahramian01-14}
{Bahramian} A.,  et~al., 2014, \mn@doi [\apj] {10.1088/0004-637X/780/2/127},
  \href {http://adsabs.harvard.edu/abs/2014ApJ...780..127B} {780, 127}

\bibitem[\protect\citeauthoryear{{Bailin} \& {Harris}}{{Bailin} \&
  {Harris}}{2009}]{bailin04-09}
{Bailin} J.,  {Harris} W.~E.,  2009, \mn@doi [\apj]
  {10.1088/0004-637X/695/2/1082}, \href
  {http://adsabs.harvard.edu/abs/2009ApJ...695.1082B} {695, 1082}

\bibitem[\protect\citeauthoryear{{Barmby}, {Huchra}, {Brodie}, {Forbes},
  {Schroder}  \& {Grillmair}}{{Barmby} et~al.}{2000}]{barmby02-00}
{Barmby} P.,  {Huchra} J.~P.,  {Brodie} J.~P.,  {Forbes} D.~A.,  {Schroder}
  L.~L.,   {Grillmair} C.~J.,  2000, \mn@doi [\aj] {10.1086/301213}, \href
  {http://adsabs.harvard.edu/abs/2000AJ....119..727B} {119, 727}

\bibitem[\protect\citeauthoryear{{Barmby}, {McLaughlin}, {Harris}, {Harris}  \&
  {Forbes}}{{Barmby} et~al.}{2007}]{barmby06-07}
{Barmby} P.,  {McLaughlin} D.~E.,  {Harris} W.~E.,  {Harris} G.~L.~H.,
  {Forbes} D.~A.,  2007, \mn@doi [\aj] {10.1086/516777}, \href
  {http://adsabs.harvard.edu/abs/2007AJ....133.2764B} {133, 2764}

\bibitem[\protect\citeauthoryear{{Bellazzini}, {Pasquali}, {Federici},
  {Ferraro}  \& {Pecci}}{{Bellazzini} et~al.}{1995}]{bellazzini02-95}
{Bellazzini} M.,  {Pasquali} A.,  {Federici} L.,  {Ferraro} F.~R.,   {Pecci}
  F.~F.,  1995, \mn@doi [\apj] {10.1086/175208}, \href
  {http://adsabs.harvard.edu/abs/1995ApJ...439..687B} {439, 687}

\bibitem[\protect\citeauthoryear{{Bregman}, {Irwin}, {Seitzer}  \&
  {Flores}}{{Bregman} et~al.}{2006}]{bregman03-06}
{Bregman} J.~N.,  {Irwin} J.~A.,  {Seitzer} P.,   {Flores} M.,  2006, \mn@doi
  [\apj] {10.1086/500037}, \href
  {http://adsabs.harvard.edu/abs/2006ApJ...640..282B} {640, 282}

\bibitem[\protect\citeauthoryear{{Bruzual} \& {Charlot}}{{Bruzual} \&
  {Charlot}}{2003}]{bruzual10-03}
{Bruzual} G.,  {Charlot} S.,  2003, \mn@doi [\mnras]
  {10.1046/j.1365-8711.2003.06897.x}, \href
  {http://adsabs.harvard.edu/abs/2003MNRAS.344.1000B} {344, 1000}

\bibitem[\protect\citeauthoryear{{Cardelli}, {Clayton}  \& {Mathis}}{{Cardelli}
  et~al.}{1989}]{cardelli10-89}
{Cardelli} J.~A.,  {Clayton} G.~C.,   {Mathis} J.~S.,  1989, \mn@doi [\apj]
  {10.1086/167900}, \href {http://adsabs.harvard.edu/abs/1989ApJ...345..245C}
  {345, 245}

\bibitem[\protect\citeauthoryear{{Carretta}, {Bragaglia}, {Gratton}, {D'Orazi}
  \& {Lucatello}}{{Carretta} et~al.}{2009}]{carretta12-09}
{Carretta} E.,  {Bragaglia} A.,  {Gratton} R.,  {D'Orazi} V.,   {Lucatello} S.,
   2009, \mn@doi [\aap] {10.1051/0004-6361/200913003}, \href
  {http://adsabs.harvard.edu/abs/2009A%26A...508..695C} {508, 695}


\bibitem[\protect\citeauthoryear{{Chabrier}}{{Chabrier}}{2003}]{chabrier07-03}
{Chabrier} G.,  2003, \mn@doi [\pasp] {10.1086/376392}, \href
  {http://adsabs.harvard.edu/abs/2003PASP..115..763C} {115, 763}

\bibitem[\protect\citeauthoryear{{Chomiuk}, {Strader}, {Maccarone},
  {Miller-Jones}, {Heinke}, {Noyola}, {Seth}  \& {Ransom}}{{Chomiuk}
  et~al.}{2013}]{chomiuk11-13}
{Chomiuk} L.,  {Strader} J.,  {Maccarone} T.~J.,  {Miller-Jones} J.~C.~A.,
  {Heinke} C.,  {Noyola} E.,  {Seth} A.~C.,   {Ransom} S.,  2013, \mn@doi
  [\apj] {10.1088/0004-637X/777/1/69}, \href
  {http://adsabs.harvard.edu/abs/2013ApJ...777...69C} {777, 69}

\bibitem[\protect\citeauthoryear{{Clark}}{{Clark}}{1975}]{clark08-75}
{Clark} G.~W.,  1975, \mn@doi [\apjl] {10.1086/181869}, \href
  {http://adsabs.harvard.edu/abs/1975ApJ...199L.143C} {199, L143}

\bibitem[\protect\citeauthoryear{{Dotter}, {Chaboyer}, {Jevremovi{\'c}},
  {Baron}, {Ferguson}, {Sarajedini}  \& {Anderson}}{{Dotter}
  et~al.}{2007}]{dotter07-07}
{Dotter} A.,  {Chaboyer} B.,  {Jevremovi{\'c}} D.,  {Baron} E.,  {Ferguson}
  J.~W.,  {Sarajedini} A.,   {Anderson} J.,  2007, \mn@doi [\aj]
  {10.1086/517915}, \href {http://adsabs.harvard.edu/abs/2007AJ....134..376D}
  {134, 376}

\bibitem[\protect\citeauthoryear{{Dotter}, {Chaboyer}, {Jevremovi{\'c}},
  {Kostov}, {Baron}  \& {Ferguson}}{{Dotter} et~al.}{2008}]{dotter09-08}
{Dotter} A.,  {Chaboyer} B.,  {Jevremovi{\'c}} D.,  {Kostov} V.,  {Baron} E.,
  {Ferguson} J.~W.,  2008, \mn@doi [\apjs] {10.1086/589654}, \href
  {http://adsabs.harvard.edu/abs/2008ApJS..178...89D} {178, 89}

\bibitem[\protect\citeauthoryear{{Dotter} et~al.,}{{Dotter}
  et~al.}{2010}]{dotter01-10}
{Dotter} A.,  et~al., 2010, \mn@doi [\apj] {10.1088/0004-637X/708/1/698}, \href
  {http://adsabs.harvard.edu/abs/2010ApJ...708..698D} {708, 698}

\bibitem[\protect\citeauthoryear{{Fabian}, {Pringle}  \& {Rees}}{{Fabian}
  et~al.}{1975}]{fabian08-75}
{Fabian} A.~C.,  {Pringle} J.~E.,   {Rees} M.~J.,  1975, \mnras, \href
  {http://adsabs.harvard.edu/abs/1975MNRAS.172P..15F} {172, 15P}

\bibitem[\protect\citeauthoryear{{Fragos} et~al.,}{{Fragos}
  et~al.}{2008}]{fragos08-08}
{Fragos} T.,  et~al., 2008, \mn@doi [\apj] {10.1086/588456}, \href
  {http://adsabs.harvard.edu/abs/2008ApJ...683..346F} {683, 346}

\bibitem[\protect\citeauthoryear{{Fragos} et~al.,}{{Fragos}
  et~al.}{2009}]{fragos09-09}
{Fragos} T.,  et~al., 2009, \mn@doi [\apjl] {10.1088/0004-637X/702/2/L143},
  \href {http://adsabs.harvard.edu/abs/2009ApJ...702L.143F} {702, L143}

\bibitem[\protect\citeauthoryear{{Gehrels}}{{Gehrels}}{1986}]{gehrels04-86}
{Gehrels} N.,  1986, \mn@doi [\apj] {10.1086/164079}, \href
  {http://adsabs.harvard.edu/abs/1986ApJ...303..336G} {303, 336}

\bibitem[\protect\citeauthoryear{{Goldsbury}, {Heyl}  \& {Richer}}{{Goldsbury}
  et~al.}{2013}]{goldsbury11-13}
{Goldsbury} R.,  {Heyl} J.,   {Richer} H.,  2013, \mn@doi [\apj]
  {10.1088/0004-637X/778/1/57}, \href
  {http://adsabs.harvard.edu/abs/2013ApJ...778...57G} {778, 57}

\bibitem[\protect\citeauthoryear{{Grindlay}}{{Grindlay}}{1993}]{grindlay01-93}
{Grindlay} J.~E.,  1993, in {Smith} G.~H.,  {Brodie} J.~P.,  eds,  Astronomical
  Society of the Pacific Conference Series Vol. 48, The Globular Cluster-Galaxy
  Connection. p.~156

\bibitem[\protect\citeauthoryear{{Harris}}{{Harris}}{1996}]{harris10-96}
{Harris} W.~E.,  1996, \mn@doi [\aj] {10.1086/118116}, \href
  {http://adsabs.harvard.edu/abs/1996AJ....112.1487H} {112, 1487}

\bibitem[\protect\citeauthoryear{{Harris}}{{Harris}}{2009}]{harris09-09}
{Harris} W.~E.,  2009, \mn@doi [\apj] {10.1088/0004-637X/703/1/939}, \href
  {http://adsabs.harvard.edu/abs/2009ApJ...703..939H} {703, 939}

\bibitem[\protect\citeauthoryear{{Harris}, {Whitmore}, {Karakla}, {Oko{\'n}},
  {Baum}, {Hanes}  \& {Kavelaars}}{{Harris} et~al.}{2006}]{harris01-06}
{Harris} W.~E.,  {Whitmore} B.~C.,  {Karakla} D.,  {Oko{\'n}} W.,  {Baum}
  W.~A.,  {Hanes} D.~A.,   {Kavelaars} J.~J.,  2006, \mn@doi [\apj]
  {10.1086/498058}, \href {http://adsabs.harvard.edu/abs/2006ApJ...636...90H}
  {636, 90}

\bibitem[\protect\citeauthoryear{{Heinke}, {Grindlay}, {Lugger}, {Cohn},
  {Edmonds}, {Lloyd}  \& {Cool}}{{Heinke} et~al.}{2003}]{heinke11-03}
{Heinke} C.~O.,  {Grindlay} J.~E.,  {Lugger} P.~M.,  {Cohn} H.~N.,  {Edmonds}
  P.~D.,  {Lloyd} D.~A.,   {Cool} A.~M.,  2003, \mn@doi [\apj]
  {10.1086/378885}, \href {http://adsabs.harvard.edu/abs/2003ApJ...598..501H}
  {598, 501}

\bibitem[\protect\citeauthoryear{{Heinke}, {Grindlay}, {Edmonds}, {Cohn},
  {Lugger}, {Camilo}, {Bogdanov}  \& {Freire}}{{Heinke}
  et~al.}{2005}]{heinke06-05}
{Heinke} C.~O.,  {Grindlay} J.~E.,  {Edmonds} P.~D.,  {Cohn} H.~N.,  {Lugger}
  P.~M.,  {Camilo} F.,  {Bogdanov} S.,   {Freire} P.~C.,  2005, \mn@doi [\apj]
  {10.1086/429899}, \href {http://adsabs.harvard.edu/abs/2005ApJ...625..796H}
  {625, 796}

\bibitem[\protect\citeauthoryear{{Heinke}, {Cohn}  \& {Lugger}}{{Heinke}
  et~al.}{2009}]{heinke02-09}
{Heinke} C.~O.,  {Cohn} H.~N.,   {Lugger} P.~M.,  2009, \mn@doi [\apj]
  {10.1088/0004-637X/692/1/584}, \href
  {http://adsabs.harvard.edu/abs/2009ApJ...692..584H} {692, 584}

\bibitem[\protect\citeauthoryear{{Humphrey} \& {Buote}}{{Humphrey} \&
  {Buote}}{2008}]{humphrey12-08}
{Humphrey} P.~J.,  {Buote} D.~A.,  2008, \mn@doi [\apj] {10.1086/592590}, \href
  {http://adsabs.harvard.edu/abs/2008ApJ...689..983H} {689, 983}

\bibitem[\protect\citeauthoryear{{Ivanova} \& {Kalogera}}{{Ivanova} \&
  {Kalogera}}{2006}]{ivanova01-06}
{Ivanova} N.,  {Kalogera} V.,  2006, \mn@doi [\apj] {10.1086/498059}, \href
  {http://adsabs.harvard.edu/abs/2006ApJ...636..985I} {636, 985}

\bibitem[\protect\citeauthoryear{{Ivanova}, {Heinke}, {Rasio}, {Belczynski}  \&
  {Fregeau}}{{Ivanova} et~al.}{2008}]{ivanova05-08}
{Ivanova} N.,  {Heinke} C.~O.,  {Rasio} F.~A.,  {Belczynski} K.,   {Fregeau}
  J.~M.,  2008, \mn@doi [\mnras] {10.1111/j.1365-2966.2008.13064.x}, \href
  {http://adsabs.harvard.edu/abs/2008MNRAS.386..553I} {386, 553}

\bibitem[\protect\citeauthoryear{{Ivanova}, {Chaichenets}, {Fregeau}, {Heinke},
  {Lombardi}  \& {Woods}}{{Ivanova} et~al.}{2010}]{ivanova07-10}
{Ivanova} N.,  {Chaichenets} S.,  {Fregeau} J.,  {Heinke} C.~O.,  {Lombardi}
  Jr. J.~C.,   {Woods} T.~E.,  2010, \mn@doi [\apj]
  {10.1088/0004-637X/717/2/948}, \href
  {http://adsabs.harvard.edu/abs/2010ApJ...717..948I} {717, 948}

\bibitem[\protect\citeauthoryear{{Ivanova} et~al.,}{{Ivanova}
  et~al.}{2012}]{ivanova12-12}
{Ivanova} N.,  et~al., 2012, \mn@doi [\apjl] {10.1088/2041-8205/760/2/L24},
  \href {http://adsabs.harvard.edu/abs/2012ApJ...760L..24I} {760, L24}

\bibitem[\protect\citeauthoryear{{Jord{\'a}n}}{{Jord{\'a}n}}{2004}]{jordan10-04}
{Jord{\'a}n} A.,  2004, \mn@doi [\apjl] {10.1086/425147}, \href
  {http://adsabs.harvard.edu/abs/2004ApJ...613L.117J} {613, L117}

\bibitem[\protect\citeauthoryear{{Jord{\'a}n} et~al.,}{{Jord{\'a}n}
  et~al.}{2004}]{jordan09-04}
{Jord{\'a}n} A.,  et~al., 2004, \mn@doi [\apj] {10.1086/422545}, \href
  {http://adsabs.harvard.edu/abs/2004ApJ...613..279J} {613, 279}

\bibitem[\protect\citeauthoryear{{Jord{\'a}n} et~al.,}{{Jord{\'a}n}
  et~al.}{2005}]{jordan12-05}
{Jord{\'a}n} A.,  et~al., 2005, \mn@doi [\apj] {10.1086/497092}, \href
  {http://adsabs.harvard.edu/abs/2005ApJ...634.1002J} {634, 1002}

\bibitem[\protect\citeauthoryear{{Katz}}{{Katz}}{1975}]{katz02-75}
{Katz} J.~I.,  1975, \mn@doi [\nat] {10.1038/253698a0}, \href
  {http://adsabs.harvard.edu/abs/1975Natur.253..698K} {253, 698}

\bibitem[\protect\citeauthoryear{{Kavelaars} \& {Hanes}}{{Kavelaars} \&
  {Hanes}}{1997}]{kavelaars03-97}
{Kavelaars} J.~J.,  {Hanes} D.~A.,  1997, \mn@doi [\mnras]
  {10.1093/mnras/285.4.L31}, \href
  {http://adsabs.harvard.edu/abs/1997MNRAS.285L..31K} {285, L31}

\bibitem[\protect\citeauthoryear{{Kim}, {Kim}, {Fabbiano}, {Lee}, {Park},
  {Geisler}  \& {Dirsch}}{{Kim} et~al.}{2006}]{kim08-06}
{Kim} E.,  {Kim} D.-W.,  {Fabbiano} G.,  {Lee} M.~G.,  {Park} H.~S.,  {Geisler}
  D.,   {Dirsch} B.,  2006, \mn@doi [\apj] {10.1086/505261}, \href
  {http://adsabs.harvard.edu/abs/2006ApJ...647..276K} {647, 276}

\bibitem[\protect\citeauthoryear{{Kim} et~al.,}{{Kim} et~al.}{2009}]{kim09-09}
{Kim} D.-W.,  et~al., 2009, \mn@doi [\apj] {10.1088/0004-637X/703/1/829}, \href
  {http://adsabs.harvard.edu/abs/2009ApJ...703..829K} {703, 829}

\bibitem[\protect\citeauthoryear{{Kim}, {Fabbiano}, {Ivanova}, {Fragos},
  {Jord{\'a}n}, {Sivakoff}  \& {Voss}}{{Kim} et~al.}{2013}]{kim02-13}
{Kim} D.-W.,  {Fabbiano} G.,  {Ivanova} N.,  {Fragos} T.,  {Jord{\'a}n} A.,
  {Sivakoff} G.~R.,   {Voss} R.,  2013, \mn@doi [\apj]
  {10.1088/0004-637X/764/1/98}, \href
  {http://adsabs.harvard.edu/abs/2013ApJ...764...98K} {764, 98}

\bibitem[\protect\citeauthoryear{{Kundu} \& {Whitmore}}{{Kundu} \&
  {Whitmore}}{1998}]{kundu12-98}
{Kundu} A.,  {Whitmore} B.~C.,  1998, \mn@doi [\aj] {10.1086/300643}, \href
  {http://adsabs.harvard.edu/abs/1998AJ....116.2841K} {116, 2841}

\bibitem[\protect\citeauthoryear{{Kundu}, {Maccarone}  \& {Zepf}}{{Kundu}
  et~al.}{2002}]{kundu07-02}
{Kundu} A.,  {Maccarone} T.~J.,   {Zepf} S.~E.,  2002, \mn@doi [\apjl]
  {10.1086/342353}, \href {http://adsabs.harvard.edu/abs/2002ApJ...574L...5K}
  {574, L5}

\bibitem[\protect\citeauthoryear{{Kundu}, {Maccarone}  \& {Zepf}}{{Kundu}
  et~al.}{2007}]{kundu06-07}
{Kundu} A.,  {Maccarone} T.~J.,   {Zepf} S.~E.,  2007, \mn@doi [\apj]
  {10.1086/518021}, \href {http://adsabs.harvard.edu/abs/2007ApJ...662..525K}
  {662, 525}

\bibitem[\protect\citeauthoryear{{Larsen}, {Brodie}, {Huchra}, {Forbes}  \&
  {Grillmair}}{{Larsen} et~al.}{2001}]{larsen06-01}
{Larsen} S.~S.,  {Brodie} J.~P.,  {Huchra} J.~P.,  {Forbes} D.~A.,
  {Grillmair} C.~J.,  2001, \mn@doi [\aj] {10.1086/321081}, \href
  {http://adsabs.harvard.edu/abs/2001AJ....121.2974L} {121, 2974}

\bibitem[\protect\citeauthoryear{{Liu}, {van Paradijs}  \& {van den
  Heuvel}}{{Liu} et~al.}{2007}]{liu06-07}
{Liu} Q.~Z.,  {van Paradijs} J.,   {van den Heuvel} E.~P.~J.,  2007, VizieR
  Online Data Catalog, \href
  {http://adsabs.harvard.edu/abs/2007yCat..34690807L} {346, 90807}

\bibitem[\protect\citeauthoryear{{Maccarone}, {Kundu}  \& {Zepf}}{{Maccarone}
  et~al.}{2004}]{maccarone05-04}
{Maccarone} T.~J.,  {Kundu} A.,   {Zepf} S.~E.,  2004, \mn@doi [\apj]
  {10.1086/382937}, \href {http://adsabs.harvard.edu/abs/2004ApJ...606..430M}
  {606, 430}

\bibitem[\protect\citeauthoryear{{Mamajek}}{{Mamajek}}{2012}]{mamajek08-12}
{Mamajek} E.~E.,  2012, \mn@doi [\apjl] {10.1088/2041-8205/754/2/L20}, \href
  {http://adsabs.harvard.edu/abs/2012ApJ...754L..20M} {754, L20}

\bibitem[\protect\citeauthoryear{{Maxwell}, {Lugger}, {Cohn}, {Heinke},
  {Grindlay}, {Budac}, {Drukier}  \& {Bailyn}}{{Maxwell}
  et~al.}{2012}]{maxwell09-12}
{Maxwell} J.~E.,  {Lugger} P.~M.,  {Cohn} H.~N.,  {Heinke} C.~O.,  {Grindlay}
  J.~E.,  {Budac} S.~A.,  {Drukier} G.~A.,   {Bailyn} C.~D.,  2012, \mn@doi
  [\apj] {10.1088/0004-637X/756/2/147}, \href
  {http://adsabs.harvard.edu/abs/2012ApJ...756..147M} {756, 147}

\bibitem[\protect\citeauthoryear{{McLaughlin} \& {van der Marel}}{{McLaughlin}
  \& {van der Marel}}{2005}]{mclaughlin12-05}
{McLaughlin} D.~E.,  {van der Marel} R.~P.,  2005, \mn@doi [\apjs]
  {10.1086/497429}, \href {http://adsabs.harvard.edu/abs/2005ApJS..161..304M}
  {161, 304}

\bibitem[\protect\citeauthoryear{{Mieske} et~al.,}{{Mieske}
  et~al.}{2006}]{mieske12-06}
{Mieske} S.,  et~al., 2006, \mn@doi [\apj] {10.1086/508986}, \href
  {http://adsabs.harvard.edu/abs/2006ApJ...653..193M} {653, 193}

\bibitem[\protect\citeauthoryear{{Miller-Jones} et~al.,}{{Miller-Jones}
  et~al.}{2015}]{miller-jones11-15}
{Miller-Jones} J.~C.~A.,  et~al., 2015, \mn@doi [\mnras]
  {10.1093/mnras/stv1869}, \href
  {http://adsabs.harvard.edu/abs/2015MNRAS.453.3919M} {453, 3919}

\bibitem[\protect\citeauthoryear{{Mineo} et~al.,}{{Mineo}
  et~al.}{2014}]{mineo01-14}
{Mineo} S.,  et~al., 2014, \mn@doi [\apj] {10.1088/0004-637X/780/2/132}, \href
  {http://adsabs.harvard.edu/abs/2014ApJ...780..132M} {780, 132}

\bibitem[\protect\citeauthoryear{{Miocchi} et~al.,}{{Miocchi}
  et~al.}{2013}]{miocchi09-13}
{Miocchi} P.,  et~al., 2013, \mn@doi [\apj] {10.1088/0004-637X/774/2/151},
  \href {http://adsabs.harvard.edu/abs/2013ApJ...774..151M} {774, 151}

\bibitem[\protect\citeauthoryear{{Nataf}, {Gould}, {Pinsonneault}  \&
  {Udalski}}{{Nataf} et~al.}{2013}]{nataf04-13}
{Nataf} D.~M.,  {Gould} A.~P.,  {Pinsonneault} M.~H.,   {Udalski} A.,  2013,
  \mn@doi [\apj] {10.1088/0004-637X/766/2/77}, \href
  {http://adsabs.harvard.edu/abs/2013ApJ...766...77N} {766, 77}

\bibitem[\protect\citeauthoryear{{Paolillo}, {Puzia}, {Goudfrooij}, {Zepf},
  {Maccarone}, {Kundu}, {Fabbiano}  \& {Angelini}}{{Paolillo}
  et~al.}{2011}]{paolillo08-11}
{Paolillo} M.,  {Puzia} T.~H.,  {Goudfrooij} P.,  {Zepf} S.~E.,  {Maccarone}
  T.~J.,  {Kundu} A.,  {Fabbiano} G.,   {Angelini} L.,  2011, \mn@doi [\apj]
  {10.1088/0004-637X/736/2/90}, \href
  {http://adsabs.harvard.edu/abs/2011ApJ...736...90P} {736, 90}

\bibitem[\protect\citeauthoryear{{Peacock}, {Maccarone}, {Kundu}  \&
  {Zepf}}{{Peacock} et~al.}{2010}]{peacock10-10}
{Peacock} M.~B.,  {Maccarone} T.~J.,  {Kundu} A.,   {Zepf} S.~E.,  2010,
  \mn@doi [\mnras] {10.1111/j.1365-2966.2010.17119.x}, \href
  {http://adsabs.harvard.edu/abs/2010MNRAS.407.2611P} {407, 2611}

\bibitem[\protect\citeauthoryear{{Peng} et~al.,}{{Peng}
  et~al.}{2006}]{peng03-06}
{Peng} E.~W.,  et~al., 2006, \mn@doi [\apj] {10.1086/498210}, \href
  {http://adsabs.harvard.edu/abs/2006ApJ...639...95P} {639, 95}

\bibitem[\protect\citeauthoryear{{Piotto} et~al.,}{{Piotto}
  et~al.}{2002}]{piotto09-02}
{Piotto} G.,  et~al., 2002, \mn@doi [\aap] {10.1051/0004-6361:20020820}, \href
  {http://adsabs.harvard.edu/abs/2002A%26A...391..945P} {391, 945}


\bibitem[\protect\citeauthoryear{{Pooley} et~al.,}{{Pooley}
  et~al.}{2003}]{pooley07-03}
{Pooley} D.,  et~al., 2003, \mn@doi [\apjl] {10.1086/377074}, \href
  {http://adsabs.harvard.edu/abs/2003ApJ...591L.131P} {591, L131}

\bibitem[\protect\citeauthoryear{{Robin}, {Reyl{\'e}}, {Derri{\`e}re}  \&
  {Picaud}}{{Robin} et~al.}{2003}]{robin10-03}
{Robin} A.~C.,  {Reyl{\'e}} C.,  {Derri{\`e}re} S.,   {Picaud} S.,  2003,
  \mn@doi [\aap] {10.1051/0004-6361:20031117}, \href
  {http://adsabs.harvard.edu/abs/2003A%26A...409..523R} {409, 523}


\bibitem[\protect\citeauthoryear{{Sarajedini} et~al.,}{{Sarajedini}
  et~al.}{2007}]{sarajedini04-07}
{Sarajedini} A.,  et~al., 2007, \mn@doi [\aj] {10.1086/511979}, \href
  {http://adsabs.harvard.edu/abs/2007AJ....133.1658S} {133, 1658}

\bibitem[\protect\citeauthoryear{{Sarazin}, {Kundu}, {Irwin}, {Sivakoff},
  {Blanton}  \& {Randall}}{{Sarazin} et~al.}{2003}]{sarazin10-03}
{Sarazin} C.~L.,  {Kundu} A.,  {Irwin} J.~A.,  {Sivakoff} G.~R.,  {Blanton}
  E.~L.,   {Randall} S.~W.,  2003, \mn@doi [\apj] {10.1086/377467}, \href
  {http://adsabs.harvard.edu/abs/2003ApJ...595..743S} {595, 743}

\bibitem[\protect\citeauthoryear{{Schlafly} \& {Finkbeiner}}{{Schlafly} \&
  {Finkbeiner}}{2011}]{schlafly08-11}
{Schlafly} E.~F.,  {Finkbeiner} D.~P.,  2011, \mn@doi [\apj]
  {10.1088/0004-637X/737/2/103}, \href
  {http://adsabs.harvard.edu/abs/2011ApJ...737..103S} {737, 103}

\bibitem[\protect\citeauthoryear{{Sivakoff} et~al.,}{{Sivakoff}
  et~al.}{2007}]{sivakoff05-07}
{Sivakoff} G.~R.,  et~al., 2007, \mn@doi [\apj] {10.1086/513094}, \href
  {http://adsabs.harvard.edu/abs/2007ApJ...660.1246S} {660, 1246}

\bibitem[\protect\citeauthoryear{{Strader} \& {Smith}}{{Strader} \&
  {Smith}}{2008}]{strader11-08}
{Strader} J.,  {Smith} G.~H.,  2008, \mn@doi [\aj]
  {10.1088/0004-6256/136/5/1828}, \href
  {http://adsabs.harvard.edu/abs/2008AJ....136.1828S} {136, 1828}

\bibitem[\protect\citeauthoryear{{Strader}, {Brodie}, {Spitler}  \&
  {Beasley}}{{Strader} et~al.}{2006}]{strader12-06}
{Strader} J.,  {Brodie} J.~P.,  {Spitler} L.,   {Beasley} M.~A.,  2006, \mn@doi
  [\aj] {10.1086/509124}, \href
  {http://adsabs.harvard.edu/abs/2006AJ....132.2333S} {132, 2333}

\bibitem[\protect\citeauthoryear{{Strader}, {Chomiuk}, {Maccarone},
  {Miller-Jones}  \& {Seth}}{{Strader} et~al.}{2012}]{strader10-12}
{Strader} J.,  {Chomiuk} L.,  {Maccarone} T.~J.,  {Miller-Jones} J.~C.~A.,
  {Seth} A.~C.,  2012, \mn@doi [\nat] {10.1038/nature11490}, \href
  {http://adsabs.harvard.edu/abs/2012Natur.490...71S} {490, 71}

\bibitem[\protect\citeauthoryear{{Trudolyubov} \& {Priedhorsky}}{{Trudolyubov}
  \& {Priedhorsky}}{2004}]{trudolyubov12-04}
{Trudolyubov} S.,  {Priedhorsky} W.,  2004, \mn@doi [\apj] {10.1086/425033},
  \href {http://adsabs.harvard.edu/abs/2004ApJ...616..821T} {616, 821}

\bibitem[\protect\citeauthoryear{{VandenBerg} \& {Clem}}{{VandenBerg} \&
  {Clem}}{2003}]{vandenberg08-03}
{VandenBerg} D.~A.,  {Clem} J.~L.,  2003, \mn@doi [\aj] {10.1086/376840}, \href
  {http://adsabs.harvard.edu/abs/2003AJ....126..778V} {126, 778}

\bibitem[\protect\citeauthoryear{{Vanderbeke}, {De Propris}, {De Rijcke},
  {Baes}, {West}  \& {Blakeslee}}{{Vanderbeke} et~al.}{2015}]{vanderbeke07-15}
{Vanderbeke} J.,  {De Propris} R.,  {De Rijcke} S.,  {Baes} M.,  {West} M.~J.,
   {Blakeslee} J.~P.,  2015, \mn@doi [\mnras] {10.1093/mnras/stv850}, \href
  {http://adsabs.harvard.edu/abs/2015MNRAS.450.2692V} {450, 2692}

\bibitem[\protect\citeauthoryear{{Verbunt} \& {Lewin}}{{Verbunt} \&
  {Lewin}}{2006}]{verbunt04-06}
{Verbunt} F.,  {Lewin} W.~H.~G.,  2006, {Globular cluster X-ray sources}.
{Cambridge: Cambridge Univ. Press}, pp 341--379

\bibitem[\protect\citeauthoryear{{Vulic}, {Gallagher}  \& {Barmby}}{{Vulic}
  et~al.}{2014}]{vulic08-14}
{Vulic} N.,  {Gallagher} S.~C.,   {Barmby} P.,  2014, \mn@doi [\apj]
  {10.1088/0004-637X/790/2/136}, \href
  {http://adsabs.harvard.edu/abs/2014ApJ...790..136V} {790, 136}

\makeatother
\end{thebibliography}

\bsp

\appendix

\section{Colour-Magnitude Diagrams For ACS Globular Clusters}\label{app:cmds-acs}

\begin{figure*}
\includegraphics[width=\textwidth]{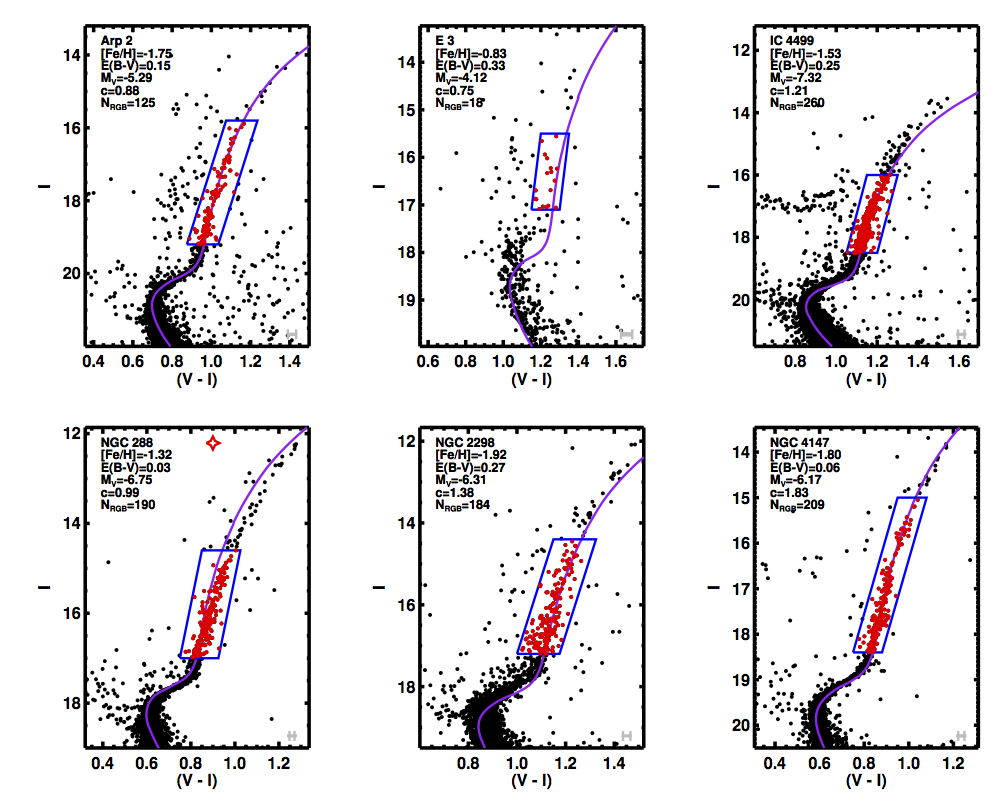}
\caption{Colour-magnitude diagrams as in Figure \ref{fig:cmdn7099} for the ACS survey GCs.}\label{fig:cmd-acsmulti}
\end{figure*}

\begin{figure*}
\includegraphics[width=\textwidth]{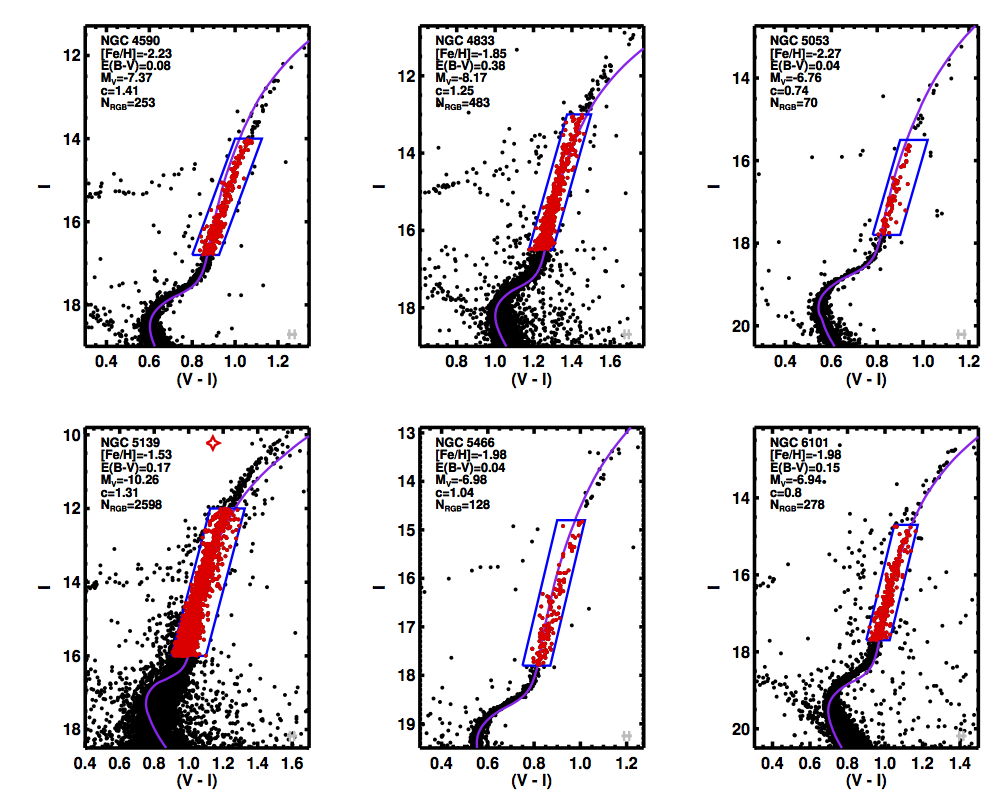}
\caption{}\label{a}
\end{figure*}

\begin{figure*}
\includegraphics[width=\textwidth]{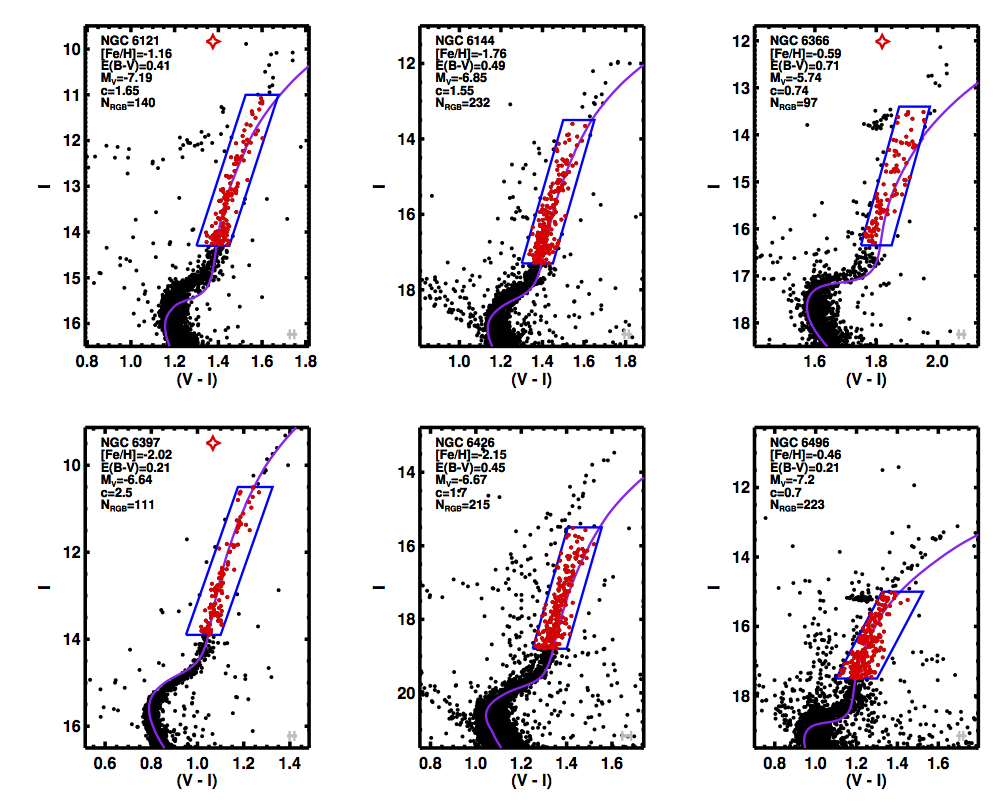}
\caption{}\label{b}
\end{figure*}

\begin{figure*}
\includegraphics[width=\textwidth]{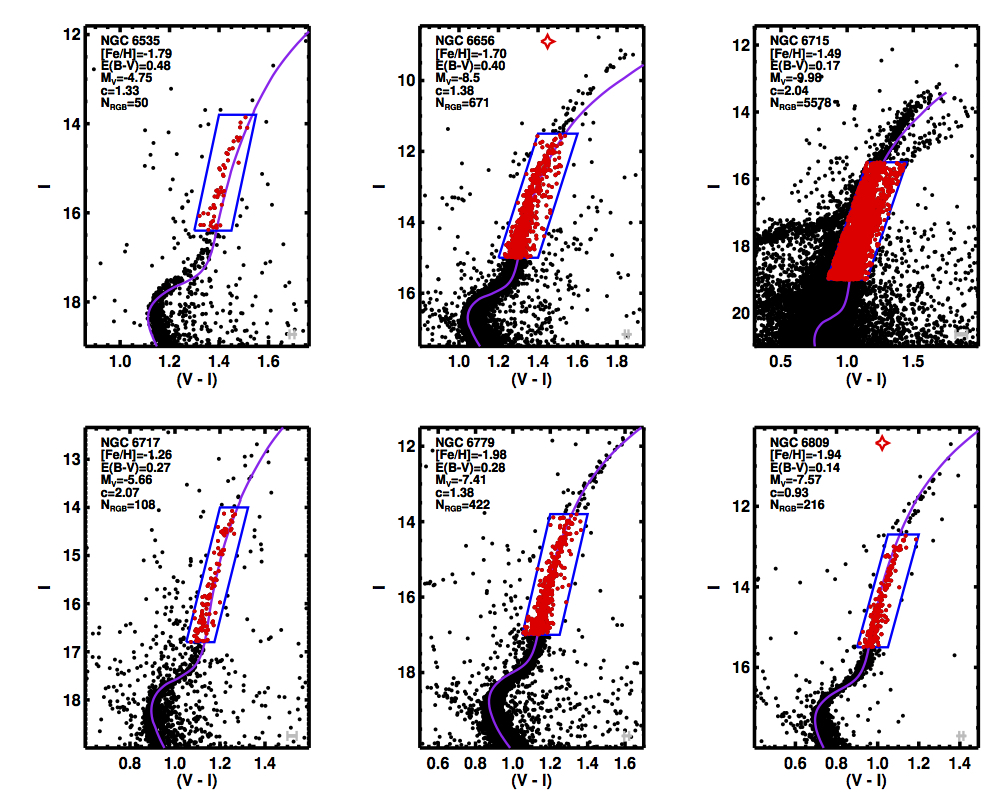}
\caption{}\label{c}
\end{figure*}

\begin{figure*}
\includegraphics[width=\textwidth]{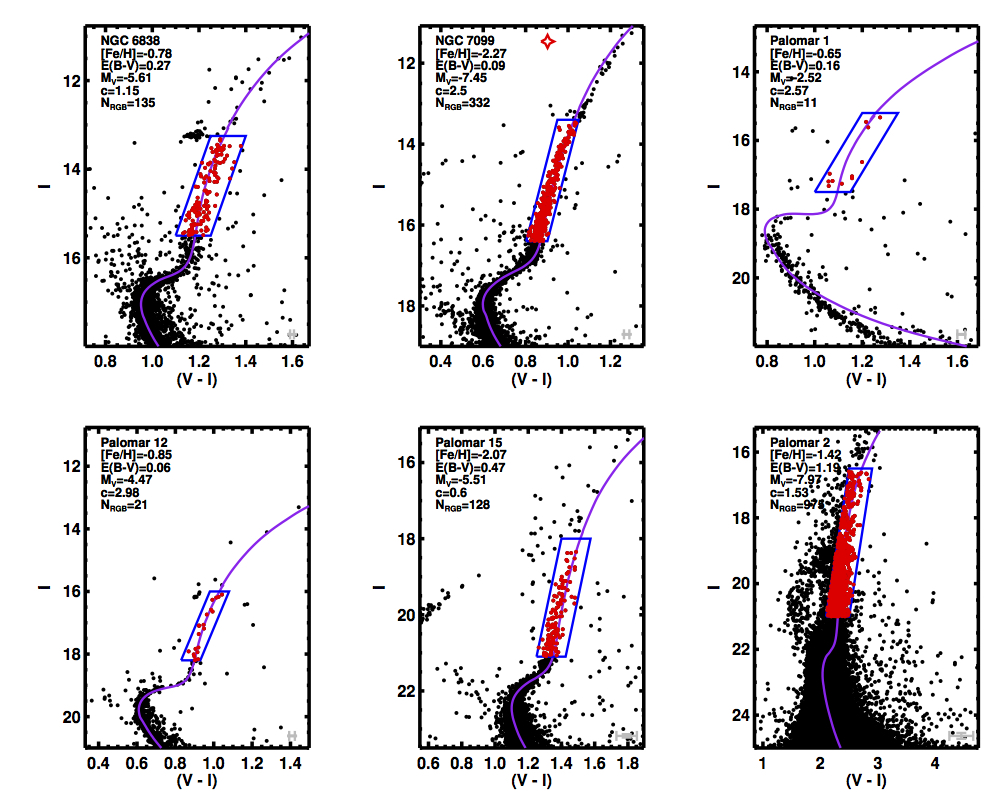}
\caption{}\label{d}
\end{figure*}

\begin{figure*}
\includegraphics[width=\textwidth]{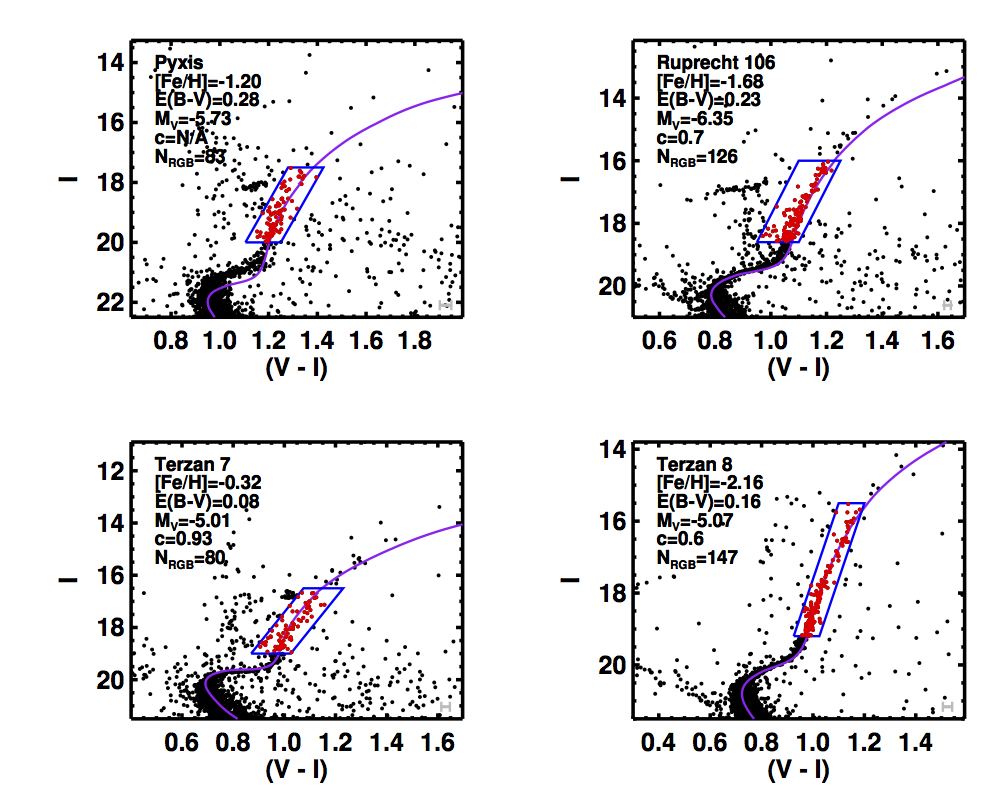}
\caption{}\label{e}
\end{figure*}

\section{Colour-Magnitude Diagrams For WFPC2 Globular Clusters}\label{app:cmds-wfpc2}

\begin{figure*}
\includegraphics[width=\textwidth]{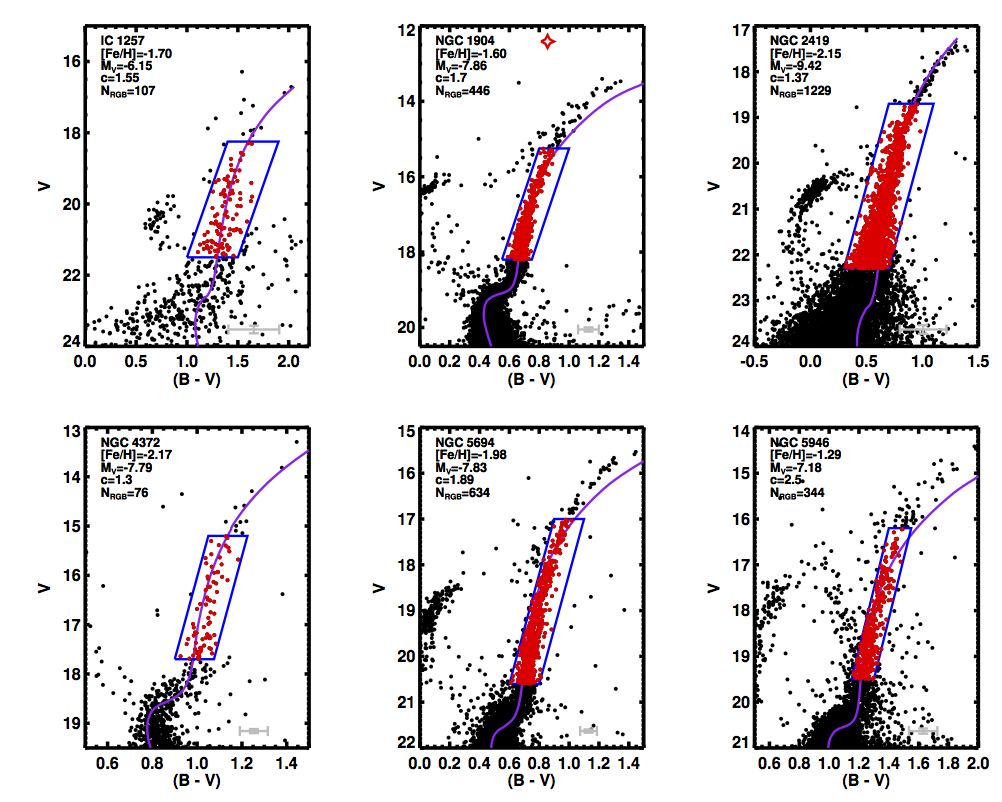}
\caption{Colour-magnitude diagrams as in Figure \ref{fig:cmdn7099} but for the WFPC2 survey GCs in the Johnson $(V,B-V)$ plane.}\label{fig:cmd-wfpc2multi}
\end{figure*}

\begin{figure*}
\includegraphics[width=\textwidth]{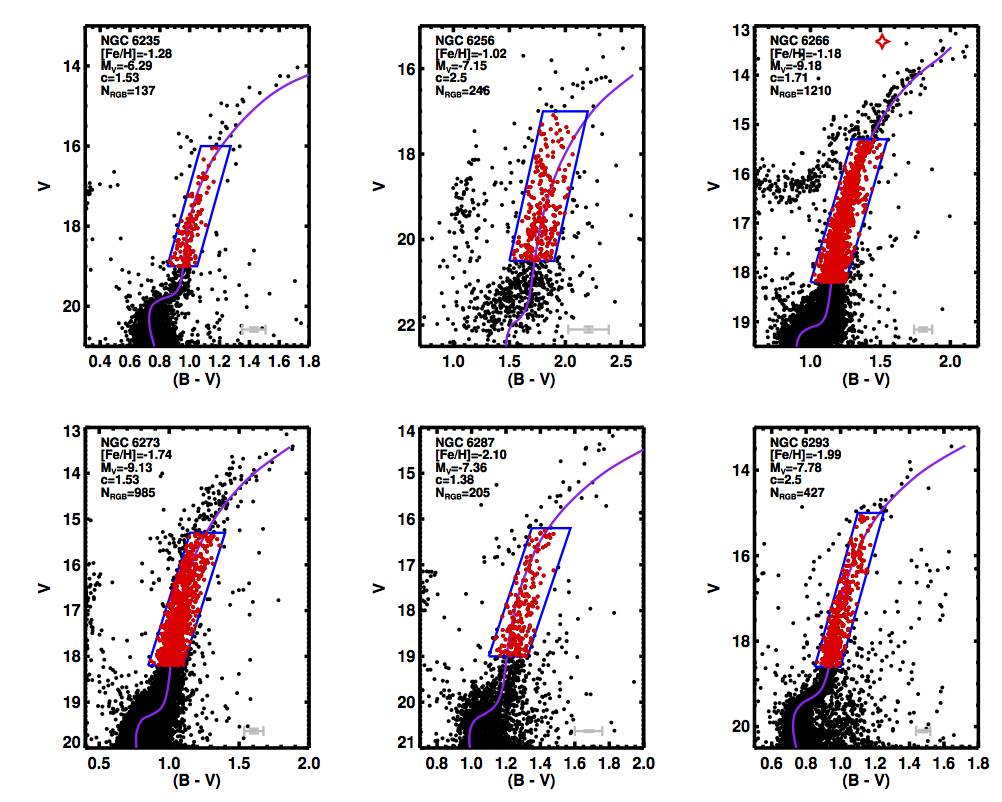}
\caption{}\label{f}
\end{figure*}

\begin{figure*}
\includegraphics[width=\textwidth]{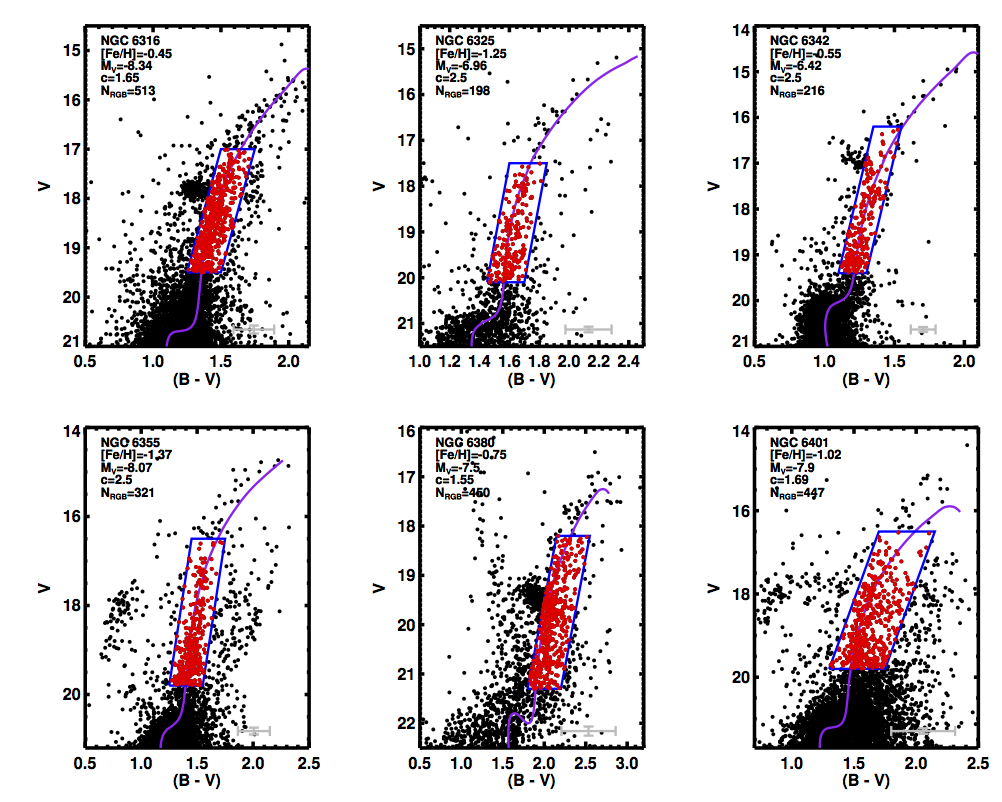}
\caption{}\label{g}
\end{figure*}

\begin{figure*}
\includegraphics[width=\textwidth]{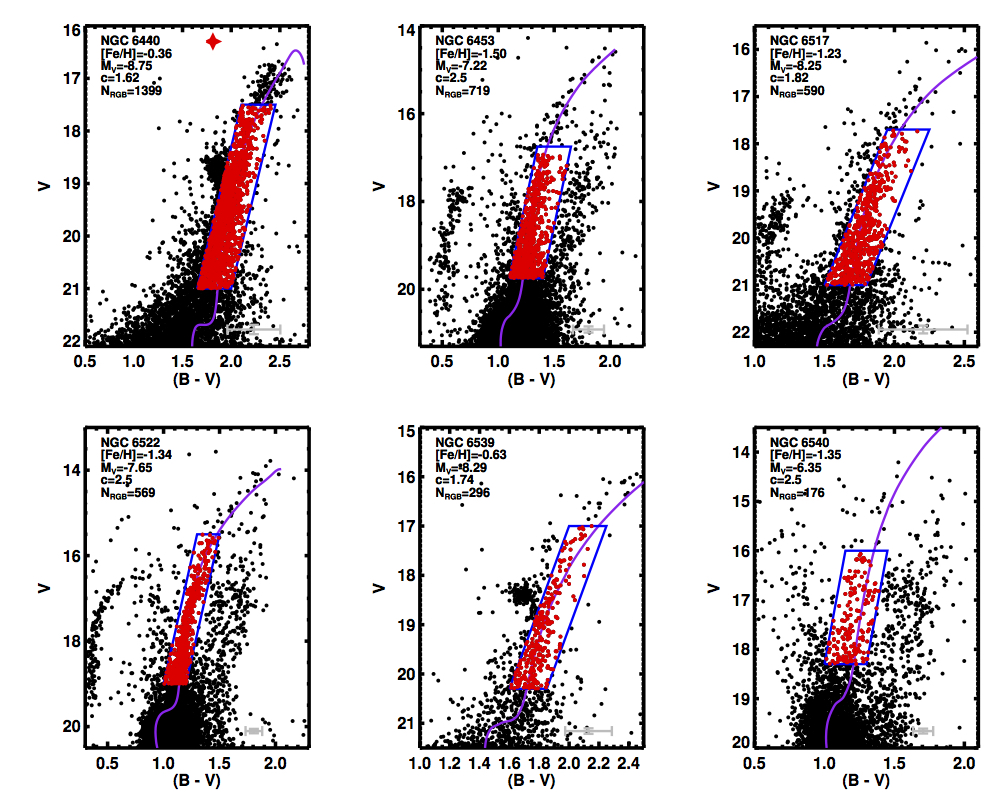}
\caption{}\label{h}
\end{figure*}

\begin{figure*}
\includegraphics[width=\textwidth]{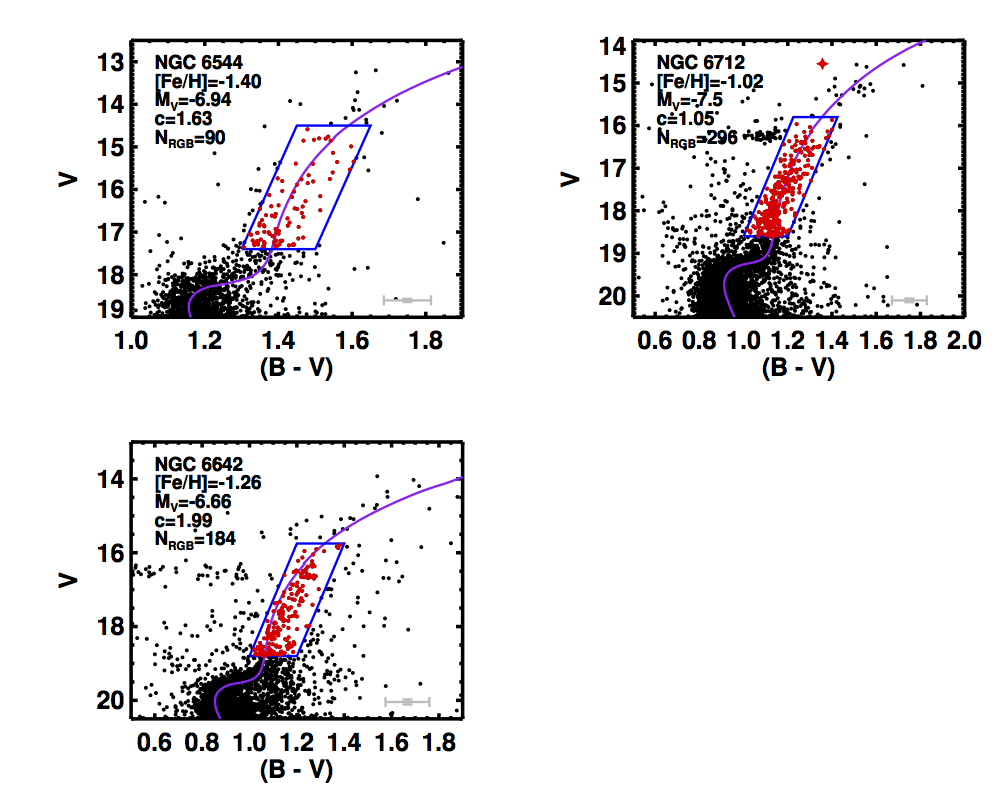}
\caption{}\label{i}
\end{figure*}

\end{document}